\documentclass[journal]{IEEEtran}
\ifCLASSINFOpdf
\else
\fi

\usepackage{times,amsmath,epsfig,subfigure,graphicx,multirow,algorithm,algorithmic,bm}
\usepackage{enumerate,paralist}
\usepackage{cite,amsfonts,dsfont,amsthm}

\begin{document}
\title{Subjective and Objective De-raining Quality Assessment Towards Authentic Rain Image}

\author{Qingbo~Wu,~Lei~Wang,~King~N.~Ngan,~Hongliang~Li,~Fanman~Meng,~and~Linfeng~Xu
\thanks{Q. Wu, L. Wang, K. N. Ngan, H. Li, F. Meng and L. Xu were with the School
of Information and Communication Engineering, University of Electronic Science and Technology of China, Chengdu,
611731 China e-mail: (qbwu@uestc.edu.cn;lwang@std.uestc.edu.cn;knngan@uestc.edu.cn;hlli@ue-stc.edu.cn;fmmeng@uestc.edu.cn;lfxu@uestc.edu.cn).}}

%



\maketitle

\begin{abstract}
Images acquired by outdoor vision systems easily suffer poor visibility and annoying interference due to the rainy weather, which brings great challenge for accurately understanding and describing the visual contents. Recent researches have devoted great efforts on the task of rain removal for improving the image visibility. However, there is very few exploration about the quality assessment of de-rained image, even it is crucial for accurately measuring the performance of various de-raining algorithms. In this paper, we first create a de-raining quality assessment (DQA) database that collects 206 authentic rain images and their de-rained versions produced by 6 representative single image rain removal algorithms. Then, a subjective study is conducted on our DQA database, which collects the subject-rated scores of all de-rained images. To quantitatively measure the quality of de-rained image with non-uniform artifacts, we propose a bi-directional feature embedding network (B-FEN) which integrates the features of global perception and local difference together. Experiments confirm that the proposed method significantly outperforms many existing universal blind image quality assessment models. To help the research towards perceptually preferred de-raining algorithm, we will publicly release our DQA database and B-FEN source code on \text{https://github.com/wqb-uestc}.
\end{abstract}

\begin{IEEEkeywords}
Single image de-raining, authentic rain image, de-raining quality assessment.
\end{IEEEkeywords}

%
\IEEEpeerreviewmaketitle

\section{Introduction}
Rainy weather often causes poor visibility and visual distraction for the image captured in outdoor scenes, which may significantly degrade the performance of various computational photography and computer vision tasks \cite{depth_deraining,deraining_GAN,rain_analysis}. In these real-world applications, image de-raining is highly desirable to achieve two objectives. First, the rain is removed as clean as possible. Second, the de-rained image is perceived as natural as possible. This is a quite challenging problem. Due to the lack of prior information for both rain and background, the single image based rain removal is highly ill-posed. Given any rain image, there are multiple alternative de-rained results, whose perceptual quality may vary significantly.

Similar to classic image restoration framework \cite{image_restoration,nonlocal_sparse_restoration}, existing algorithms typically model single image rain removal as a decomposition problem, which aims to separate contaminated image into the rain and background layers. To make it tractable, various techniques are proposed to describe the prior information for rain and background. In \cite{guided_filter,guided_filter_removal,multi-filter}, a set of guided filter based methods are proposed to model the background prior from specific rain-free or pre-processing samples. Due to the sensitivity to guidance image and parameter selection, these methods easily cause over- or under-smooth for rain image, which would damage original image structure or leave too many rain. In \cite{low_rank_model}, Chen \textit{et al.} utilize low-rank approximation to capture the rain prior. It works efficiently in estimating rain streaks, but also easily mistakes striped background, whose texture is similar to the rain streaks. To better distinguish rain streaks from background, many researchers propose to simultaneously learn the priors for rain and background layers via dictionary learning, Gaussian mixture model, deep neural network and so on \cite{removal_decomposition,DSP_rain_removal,layer_prior_removal,joint_detection_removal,CNN_rain_removal}. These data-driven methods deliver great perform for the rain images whose appearance features are covered by the training samples. But, in dealing with some rare types of rain images with respect to the training set, their performance would drop significantly as well.

In the real-world application, rain image will present great diversity due to the change of background illumination, lens speed, depth of field, and so on. Although there are multiple de-raining algorithms developed recently, they usually capture partial properties of rain images, whose performance may change significantly from one image to another one. To select optimal de-raining result for each single image and find the further direction towards universal rain removal, it has become crucial to accurately evaluate the de-rained images produced by different algorithms. However, surprisingly, there are quite rare literatures exploring the perceptual evaluation for de-raining algorithms. Existing methods typically evaluate de-raining performance on a few synthesized rain images\footnote{http://www.photoshopessentials.com/photo-effects/rain/}, whose ground-truth images (i.e., rain-free versions) are available. Then, two classic full-reference objective metrics including PSNR and SSIM \cite{SSIM} are employed for quantitative quality assessment. In comparison with the diverse authentic rain images, these synthetic data only cover very limited types of raindrops, which are far from sufficient to verify the de-raining capability in reality. Meanwhile, given a de-rained sample produced from authentic rain image, it is also challenging to accurately evaluate its performance due to the absence of ground-truth image.

\begin{figure}[t]
  \centering
  \includegraphics[width=\linewidth]{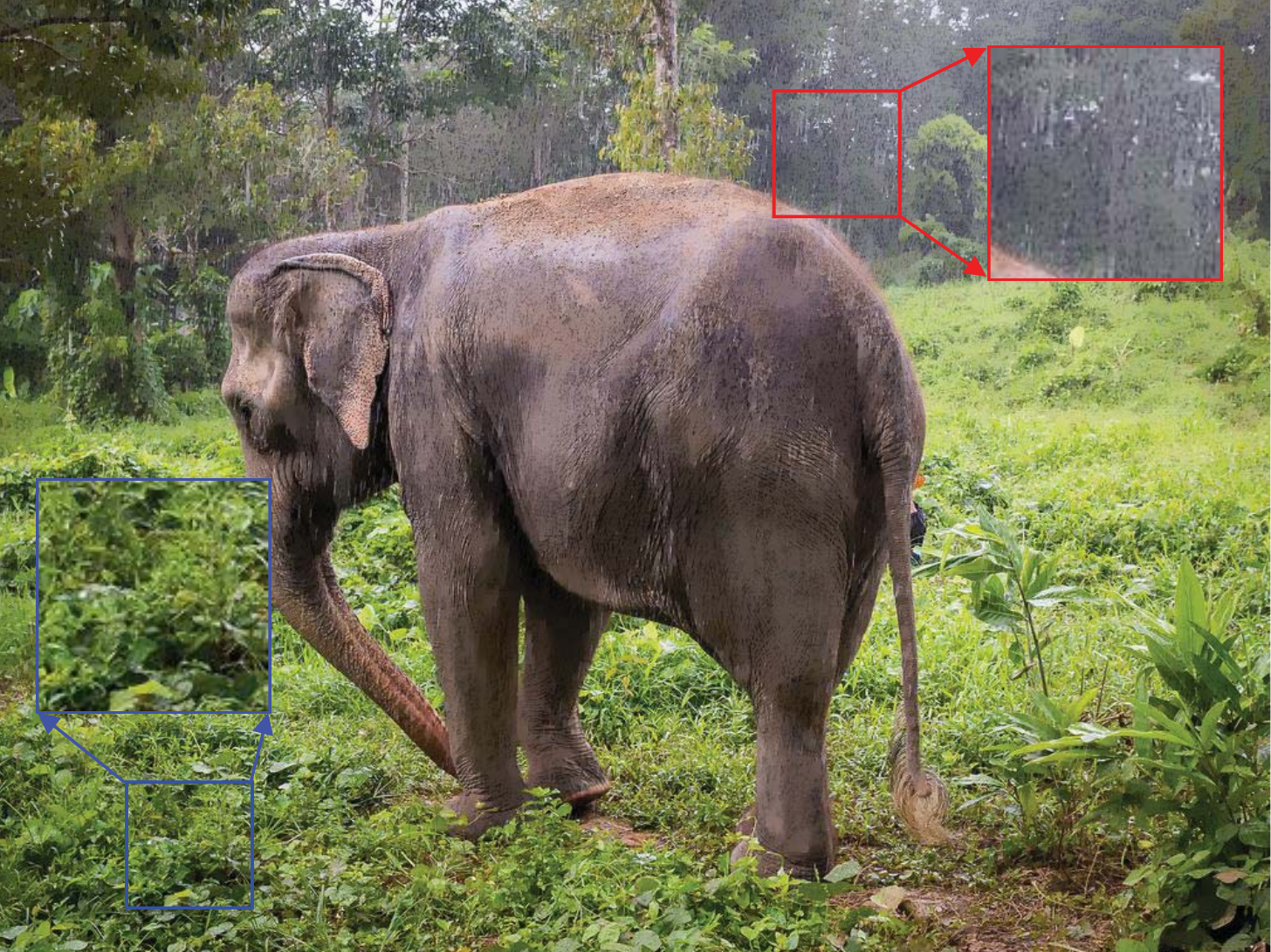}\\
  \caption{Illustration of non-uniform de-raining artifacts. This example is generated by \cite{guided_filter_removal}, where the red and blue bounding-boxes highlight the de-raining results for different local regions, respectively. It is clear that the red bounding-box presents poor quality due to the annoying holes and streaks. By contrast, the blue bounding-box presents good quality where all leaves retain natural appearance.}
  \label{fig-de-raining-artifact}
\end{figure}

To the best of our knowledge, the first exploration of de-raining quality assessment (DQA) is conducted in our previous work \cite{B-GFN}, which proposed a no-reference image quality assessment (NR-IQA) model specifically designed for the de-rained image and investigated the performance of many existing general purpose NR-IQA models \cite{LOCRUE,BIQI,TCLT,R3,DIIVINE,BRISQUE,MSGF-PR} in the DQA task. In this paper, we extend our previous exploration \cite{B-GFN} of subjective and objective DQA tasks towards the authentic rain images. More comprehensive statistical analysis is conducted for the subject-rated data and an enhanced deep feature representation is proposed to improve the performance of DQA. The detailed contributions are summarized in the following:
\begin{enumerate}
  \item IVIPC-DQA database: We create a DQA database which collects a variety of authentic rain images and their de-rained versions produced by 6 categories of single image rain removal algorithms. Then, a subjective study is conducted on the DQA database, which presents two important findings. Firstly, although existing de-raining algorithms perform well on removing synthetic rain streaks, they still hardly balance the rain removal and detail preservation towards authentic rain image. Secondly, existing general purpose NR-IQA models perform poorly in evaluating the de-rained image, whose artifacts present significantly different characteristics with respect to the traditional uniform distortions, such as, the white noise or gaussian blur.
  \item B-FEN DQA model: We propose a bi-directional feature embedding network to accurately assess the de-raining quality. The de-raining artifacts usually present different degradation degrees in different local regions as shown in Fig. \ref{fig-de-raining-artifact}, which brings great difficulty in identifying the overall quality of a de-rained image. To cope with this issue, we employ a forward branch to suppress the quality irrelevant information at the cost of dimension reduction for the feature maps. Then, a backward branch is developed to embed the low-resolution quality-aware features into the high-resolution feature maps of shallow layers. A gated fusion module is further utilized to integrate the forward and backward features together, which captures both the global perception and local difference.
\end{enumerate}

Extensive experiments on our IVIPC-DQA database demonstrate that the proposed B-FEN model significantly outperform many classic general purpose NR-IQA models and the latest deep learning based quality evaluators in the task of de-raining quality assessment.

The rest of this paper is organized as follows. In Section II, we first introduce the IVIPC-DQA database and our findings from the subjective investigation. The Section III describes the proposed B-FEN model in details, and the experimental results will be shown in Section IV. Finally, we conclude this paper in Section V.

\section{Subjective study of DQA}

\subsection{Image Collection}

\begin{figure}[t]
\centering
  \includegraphics[width=\linewidth]{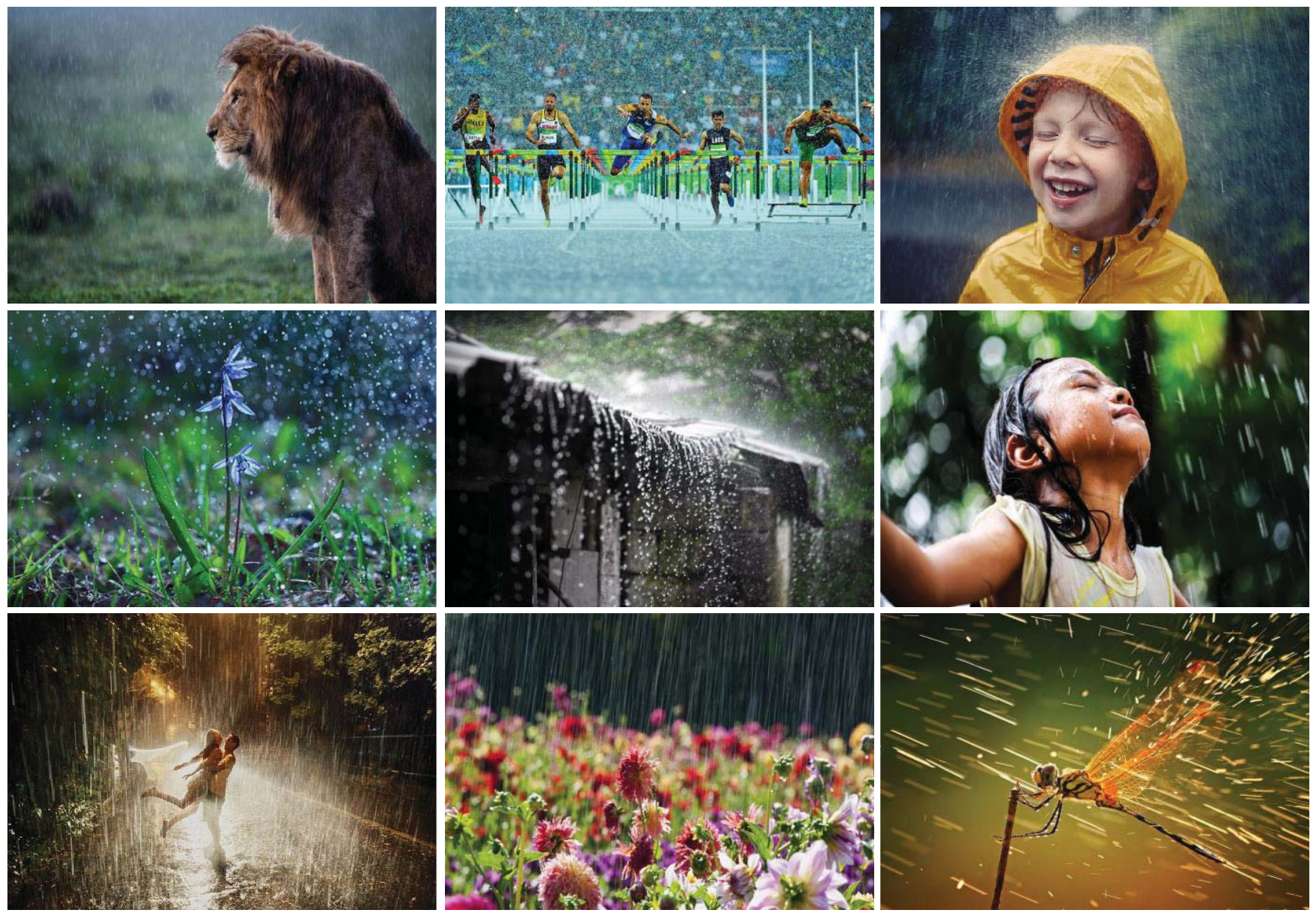}\\
  \caption{Authentic rain images collected in the DQA database.}
  \label{fig_database}
\end{figure}

In order to cover diverse rain scenes, we first collect 206 authentic rain images from Internet, which are captured under different illuminations, perspectives, lens speeds, depth of fields, and so on. Some sample images are shown in Fig. \ref{fig_database}.
In the following, we apply six representative single image rain removal algorithms to these authentic rain images, which generate totally 1236 de-rained samples. To avoid the composite distortion caused by compression, both the source and de-rained images are saved with lossless compression format.

\begin{figure}[t]
\centering
  \subfigure[Ding16 \cite{guided_filter_removal}]{
  \includegraphics[width=.45\linewidth]{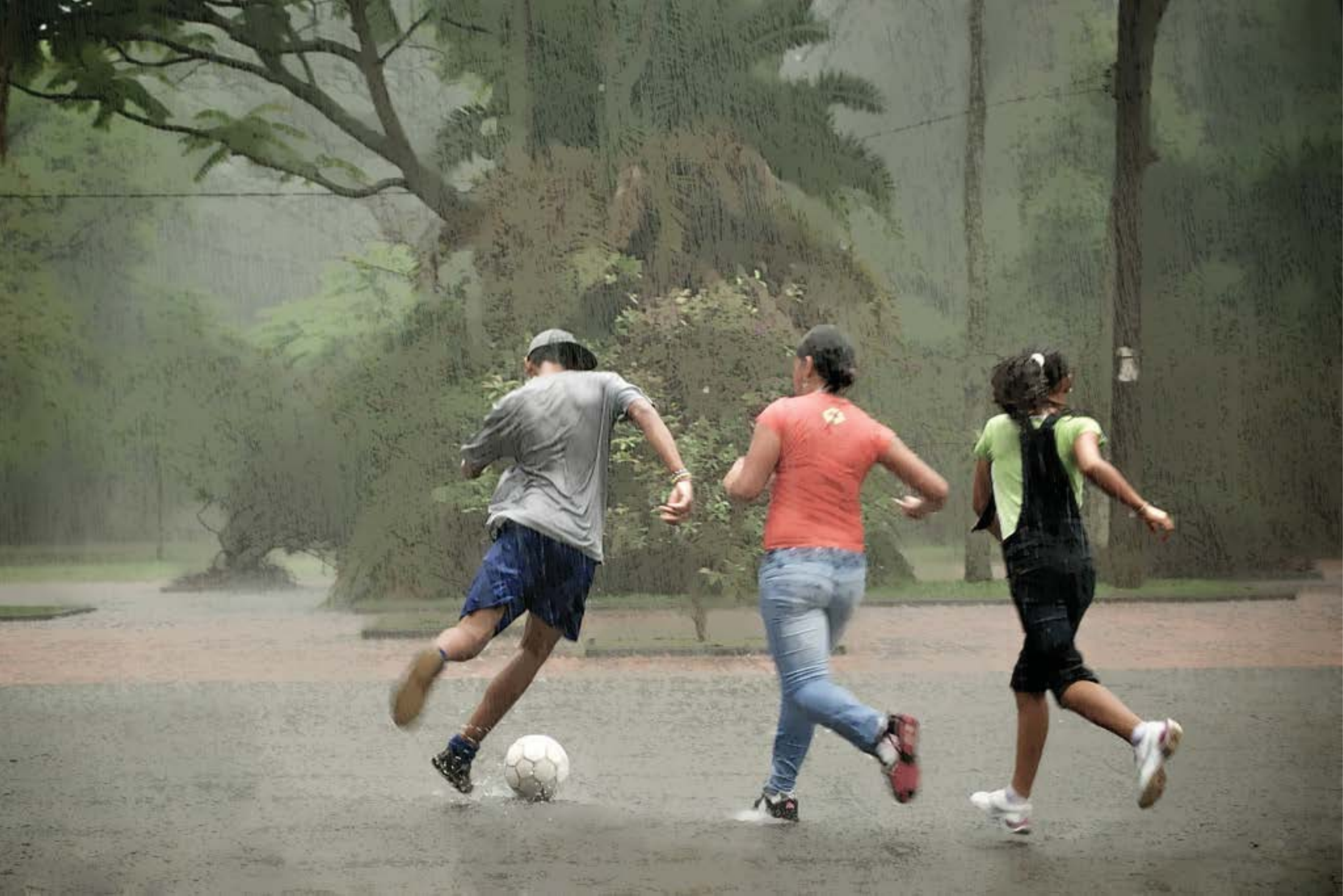}}
  \subfigure[Kang12 \cite{removal_decomposition}]{
  \includegraphics[width=.45\linewidth]{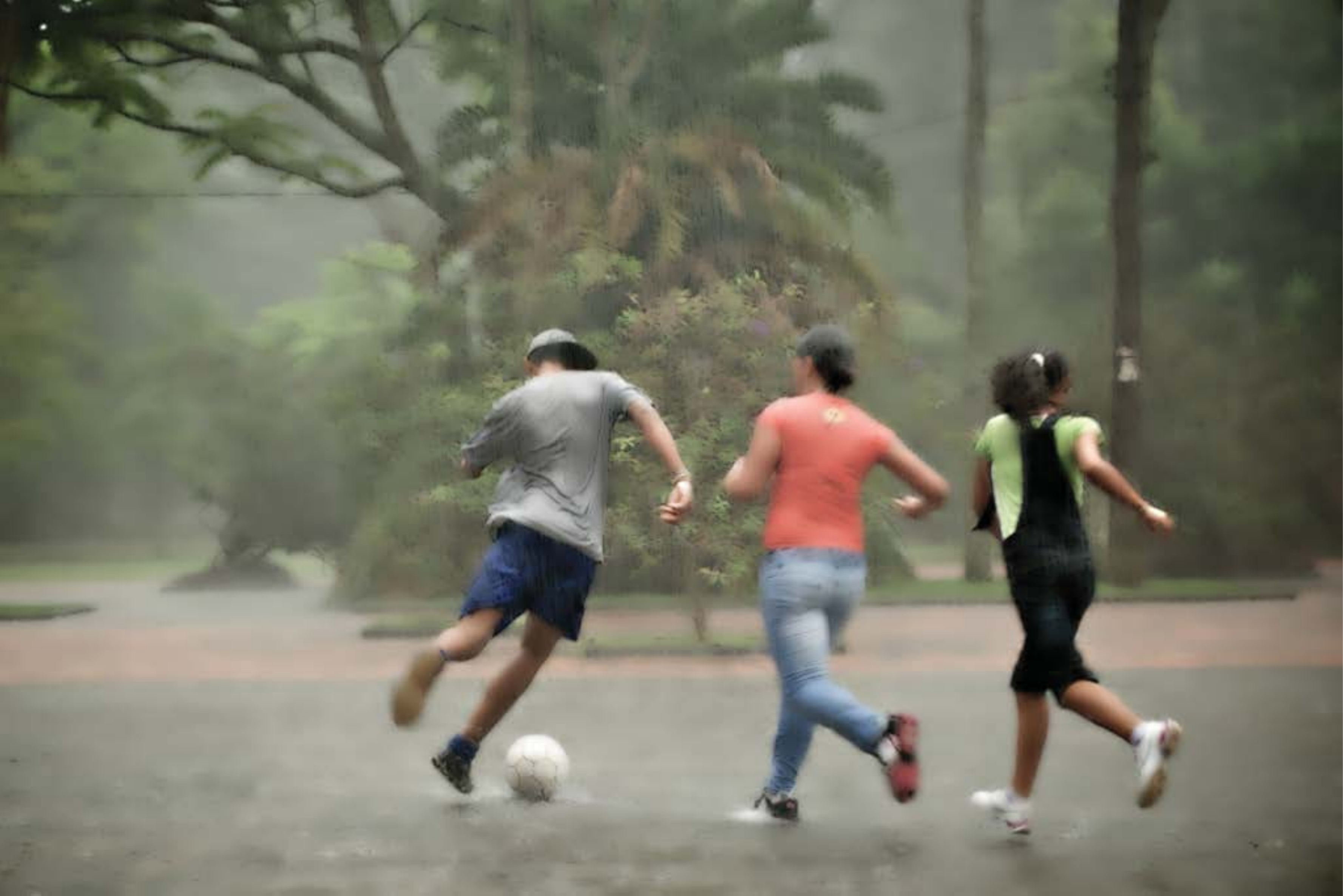}}
  \subfigure[Luo15 \cite{DSP_rain_removal}]{
  \includegraphics[width=.45\linewidth]{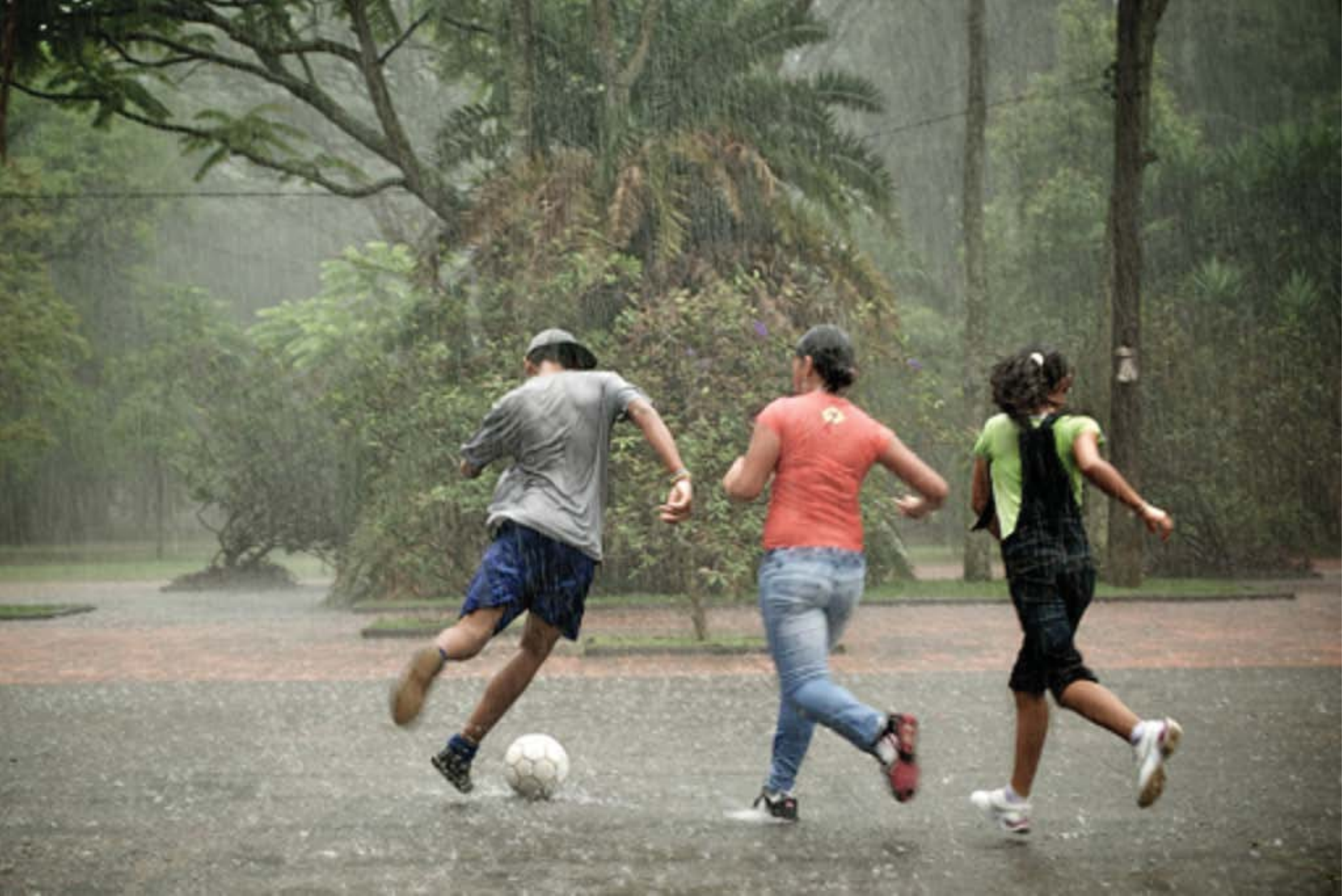}}
  \subfigure[Li16 \cite{layer_prior_removal}]{
  \includegraphics[width=.45\linewidth]{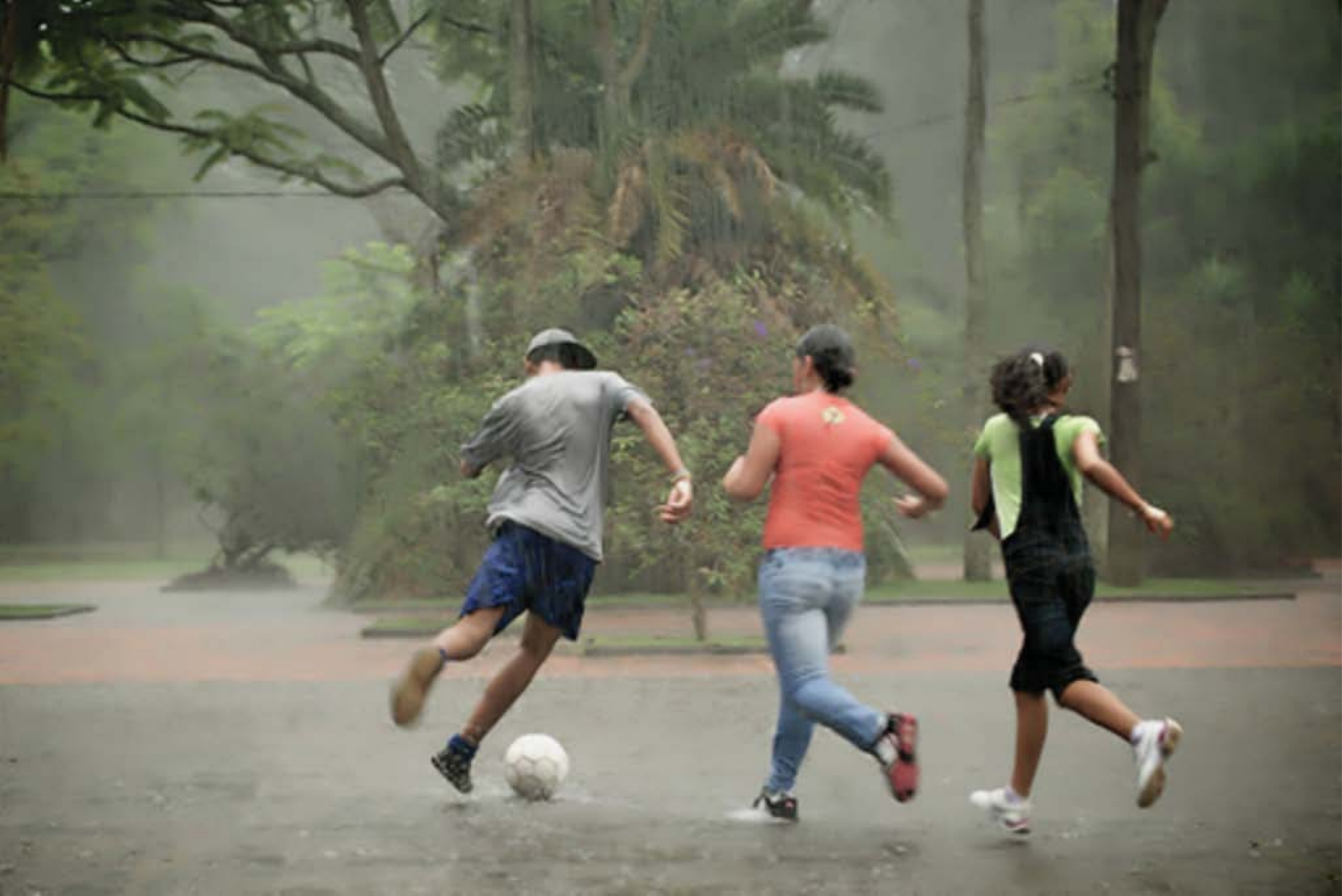}}
  \subfigure[Deng17 \cite{UGSM}]{
  \includegraphics[width=.45\linewidth]{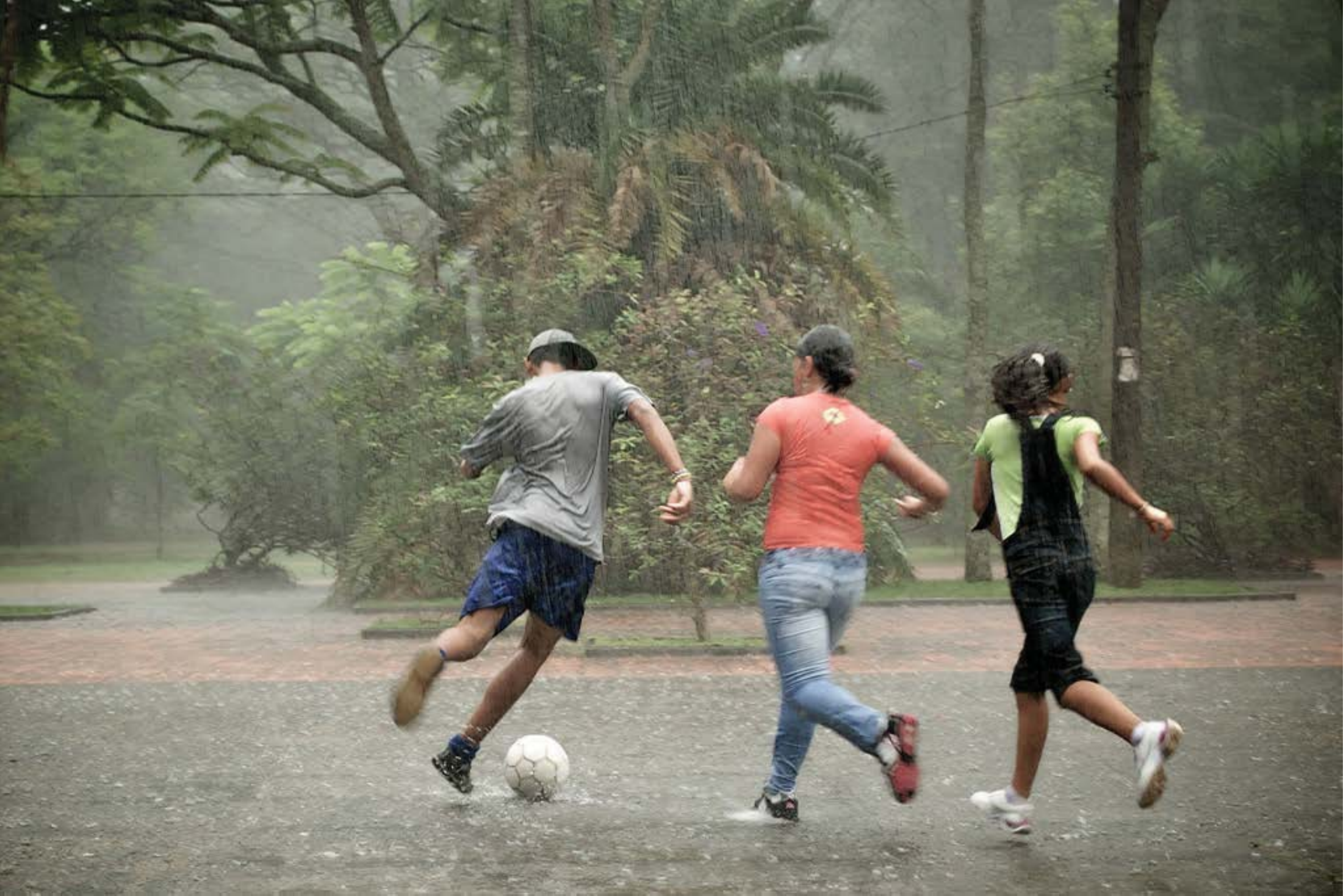}}
  \subfigure[Fu17 \cite{CNN_rain_removal}]{
  \includegraphics[width=.45\linewidth]{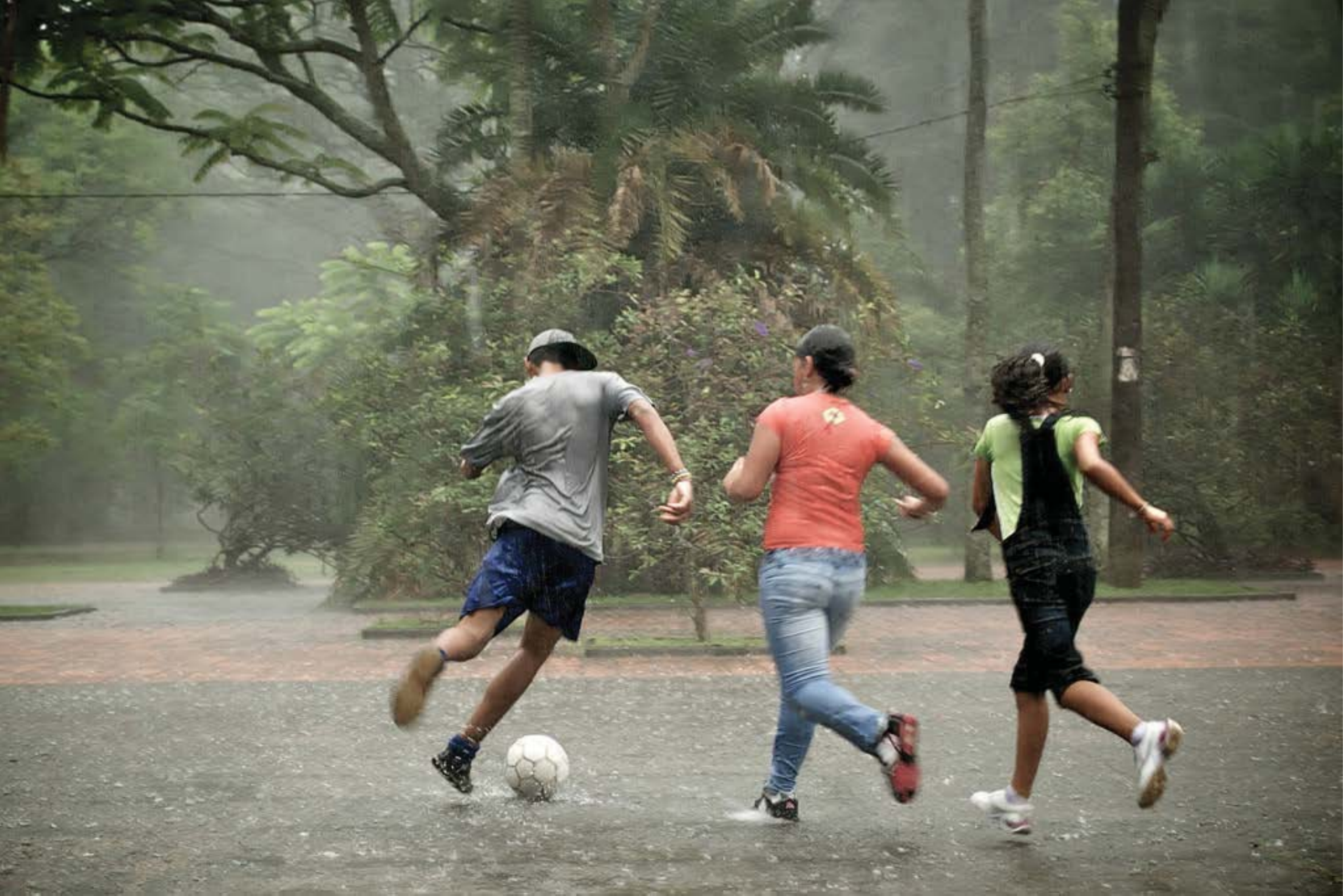}}
  \subfigure[Input]{
  \includegraphics[width=.45\linewidth]{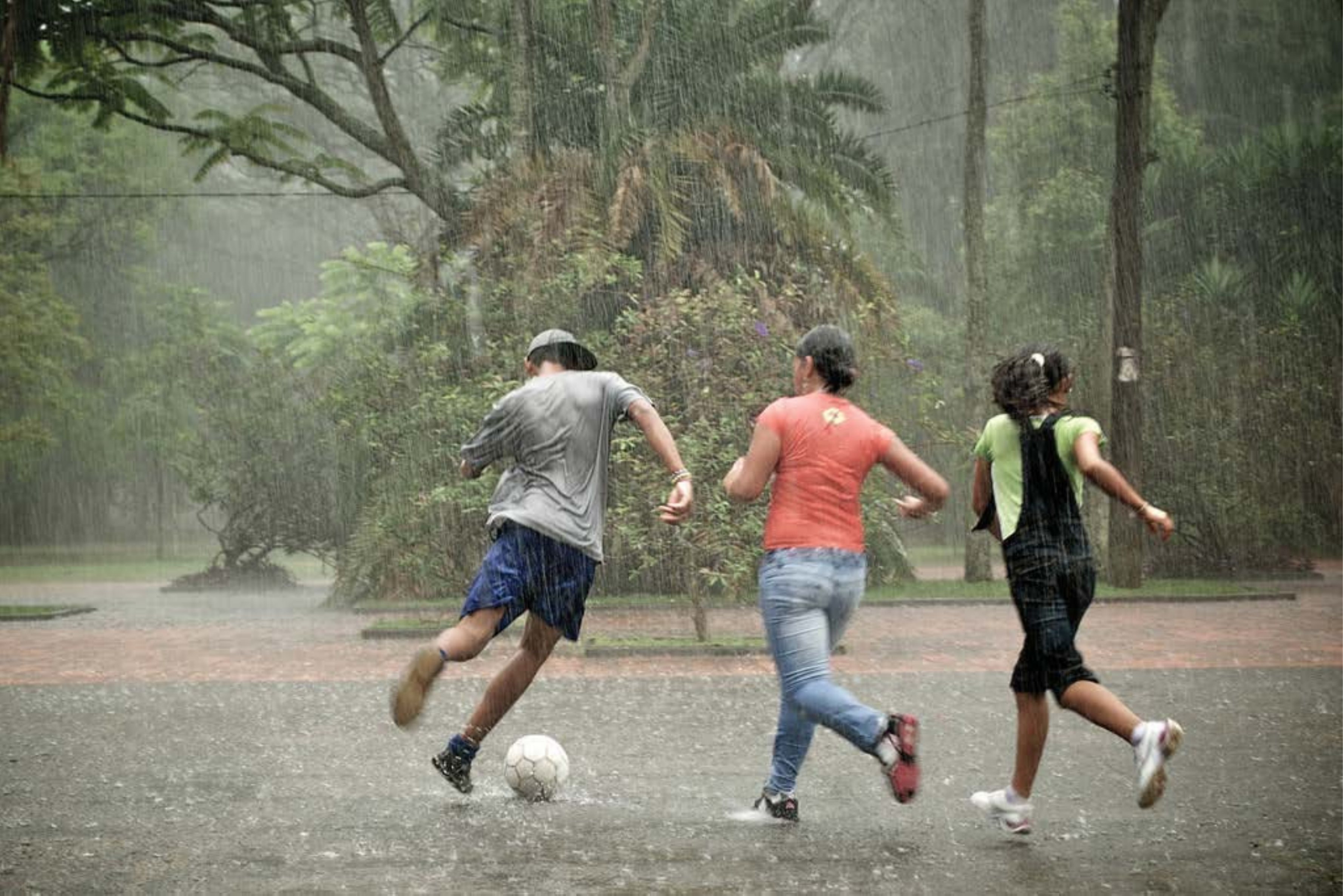}}
  \caption{Illustration of de-rained images produced by different algorithms.}
  \label{fig_derain_sample}
\end{figure}

The de-raining algorithms investigated in this paper cover a wide variety of techniques, which include the guided filter, dictionary learning, low-rank approximation, maximum posteriori, directional regularization and deep neural network. For short, we denote them by Ding16 \cite{guided_filter_removal}, Kang12 \cite{removal_decomposition}, Luo15 \cite{DSP_rain_removal}, Li16 \cite{layer_prior_removal}, Deng17 \cite{UGSM}, and Fu17 \cite{CNN_rain_removal}. In our investigation, all codes are provided by authors and the default parameters are used without additional fine-tuning procedure.
Given each authentic rain image, there are a set of six de-rained results available in our database. For illustration, an intuitive comparison between different de-raining algorithms are given in Fig. \ref{fig_derain_sample}. It is clear that different de-rained images present obviously different appearances as shown in Figs. \ref{fig_derain_sample} (a)-(f). In the following, we implement a subjective study to quantitatively evaluate these de-raining algorithms.

\subsection{Subjective Testing Method}

The subjective experiment is conducted in Intelligent Visual Information Processing and Communication (IVIPC) laboratory, UESTC\footnote{http://ivipc.uestc.edu.cn/}. All images are displayed on a 27-inch true color (32 bits) LCD monitor with the resolution of 1920$\times$1080. The viewing conditions are set by following the recommendation of ITU-R BT.500-13 \cite{BT500}. In total, there are 22 naive subjects participated in this experiment, which include 11 males and 11 females.

Specific to the rain removal task, we require all participants to rate the derained images by 5 levels which is represented via a continuous scale between 1 to 100. A lower rating score indicates a worse perceptual quality, which still retains the rain or distorts original image structure. By contrast, a higher rating score denotes a better perceptual quality, which not only removes the rain but also well preserves original image structure. The detailed description about the rating criteria is given in Table \ref{table_rating_level}.

\begin{figure}[t]
  \centering
  \includegraphics[width=\linewidth]{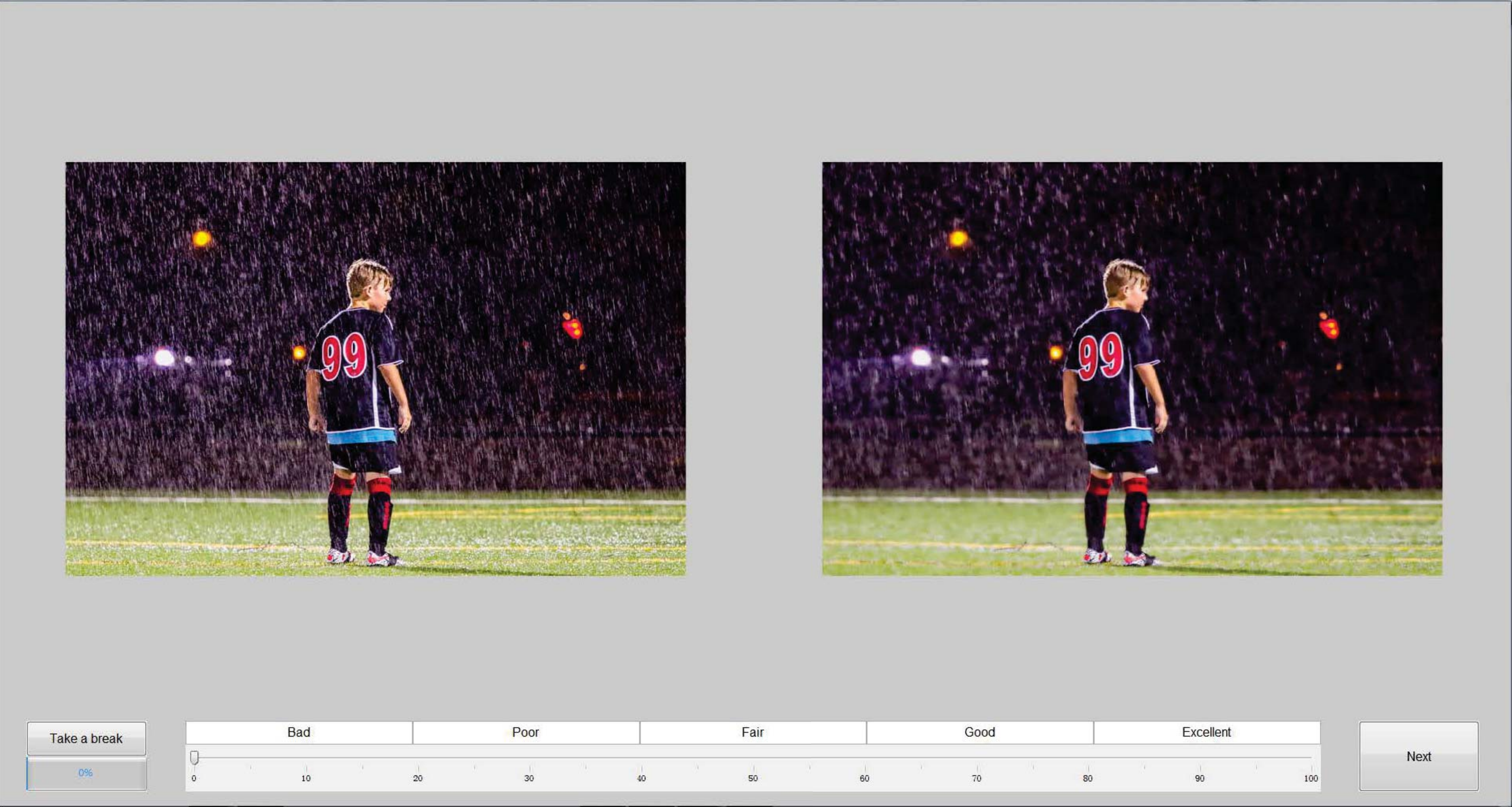}\\
  \caption{The dialogue window of our subjective experiment.}
  \label{fig_diag_window}
\end{figure}

\begin{table}[t]
  \centering
  \caption{Rating criteria for the rain removal task}
  \includegraphics[width=\linewidth]{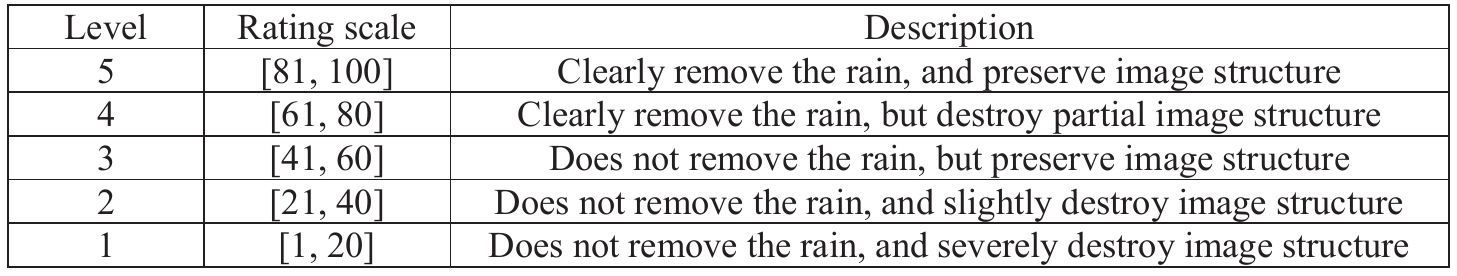}
  \label{table_rating_level}
\end{table}

Following the recommendation of ITU-T P.910 \cite{P910}, we employ the simultaneous presentation method to evaluate the de-raining performance. The reference image (i.e., rain image) and its associated de-rained version are simultaneously presented to the subject via a customized dialogue window. The reference image is always placed on the left and all subjects are aware of the relative positions of these two images. For clarity, the dialogue window of our subjective experiment is shown in Fig. \ref{fig_diag_window}. To avoid the memory effect in the human rating, we randomly show a de-rained version for the given reference image in each time, which is selected from six de-raining algorithms. For each participant, the human rating is implemented by 1236 times until all de-rained images are assigned to a corresponding rating score. Meanwhile, to reduce the influence of fatigue
effect, the duration of each rating session is limited to 30 minutes, which allows the participants to take a break after rating several pairs of images.

The raw scores collected from multiple subjects may contain a few outliers. We first clean the human rating scores via the $\beta_2$ test based outlier rejection method \cite{BT500}. Then, all raw scores are tuned into Z-scores \cite{score_range,retarget}, which is demonstrated efficient in eliminating the individual difference. Let $s_{i,j}$ denote the raw score, which is obtained from the $i$th subject in evaluating the $j$th de-rained image. Let $\bar{s}_i$ and $\sigma_i$ denote the mean score and standard deviation of the $i$th subject across all de-rained images, respectively. The Z-score $z_{i,j}$ could be computed by
\begin{equation}
z_{i,j}=\frac{s_{i,j}-\bar{s}_i}{\sigma_i}
\end{equation}

Similar to \cite{score_range,retarget}, we further rescale the Z-scores to the range of [0, 100] via a linear mapping function
\begin{equation}
\hat{z}_{i,j}=\frac{100\cdot(z_{i,j}+3)}{6}
\end{equation}
which assumes the Z-scores of each subject follow Gaussian distribution and nearly 99\% Z-scores fall into the range of [-3, 3]. Finally, the mean opinion score (MOS) of the $j$th de-rained image is computed by
\begin{equation}
m_j = \frac{\sum_{i=1}^{N_j}\hat{z}_{i,j}}{N_j}
\end{equation}
where $N_j$ is the number of valid human ratings for the $j$th de-rained image.

\subsection{Analysis of Human Ratings}

\begin{figure}[t]
  \centering
  \includegraphics[width=.95\linewidth]{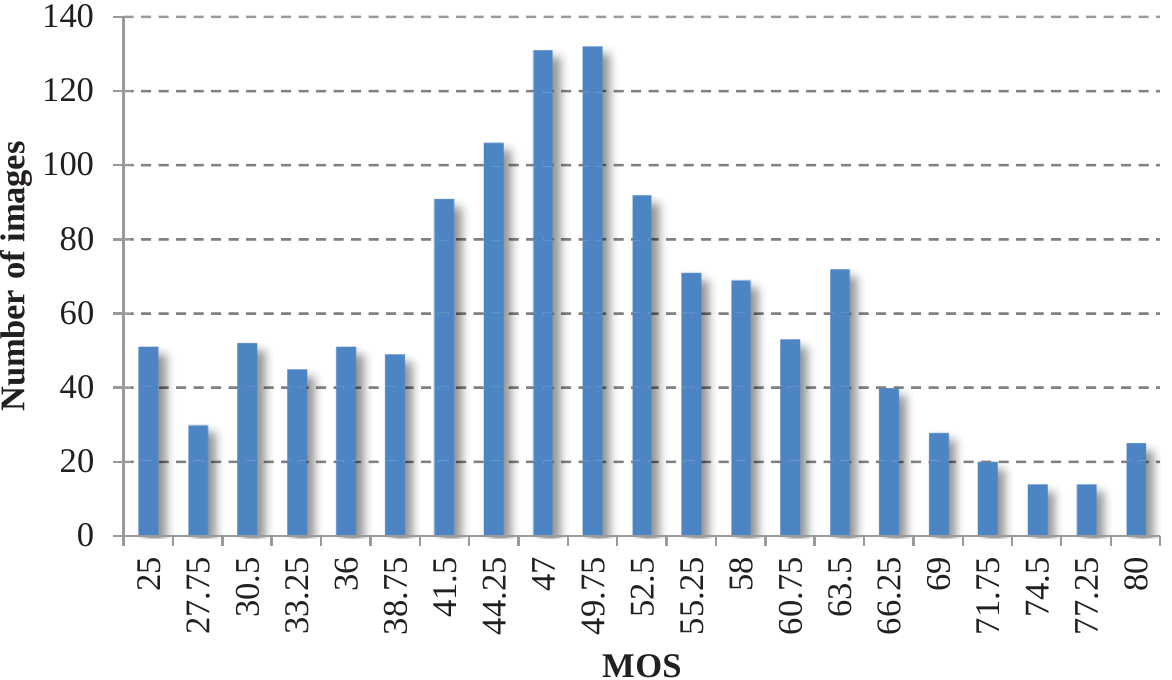}\\
  \caption{The distribution of MOS values for our DQA database.}
  \label{fig_hist_mos}
\end{figure}

To investigate the de-raining performance of existing algorithms, we first illustrate the distribution of MOS values for all de-rained images in Fig. \ref{fig_hist_mos}. It is observed that the collected perceptual qualities span a wide range from the low to high scores. Meanwhile, the distribution of our collected perceptual qualities show reasonably uniform fashion. This good separation of perceptual qualities facilitates a more reliable investigation on the perceptual characteristics of de-rained images \cite{score_range,wireless_video}.

\begin{figure}[t]
  \centering
  \includegraphics[width=.8\linewidth]{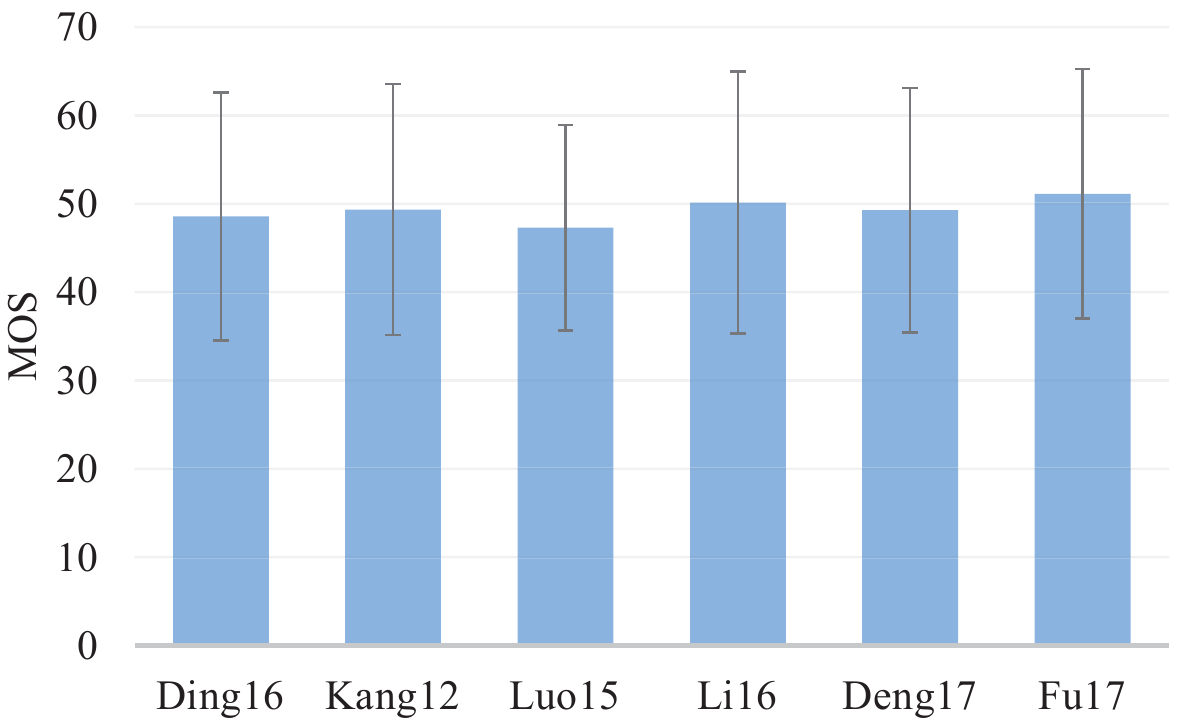}\\
  \caption{The mean and standard deviation comparison of MOS values between different de-raining algorithms.}
  \label{fig_mos_comparison}
\end{figure}

\begin{table}[t]
  \centering
  \caption{Results of onesided \textit{t-test} between the MOS values of different de-raining algorithms. A value of ``1''/``0''/``-1'' indicates that the row algorithm is statistically superior/equivalent/inferior to the column algorithm}
  \includegraphics[width=\linewidth]{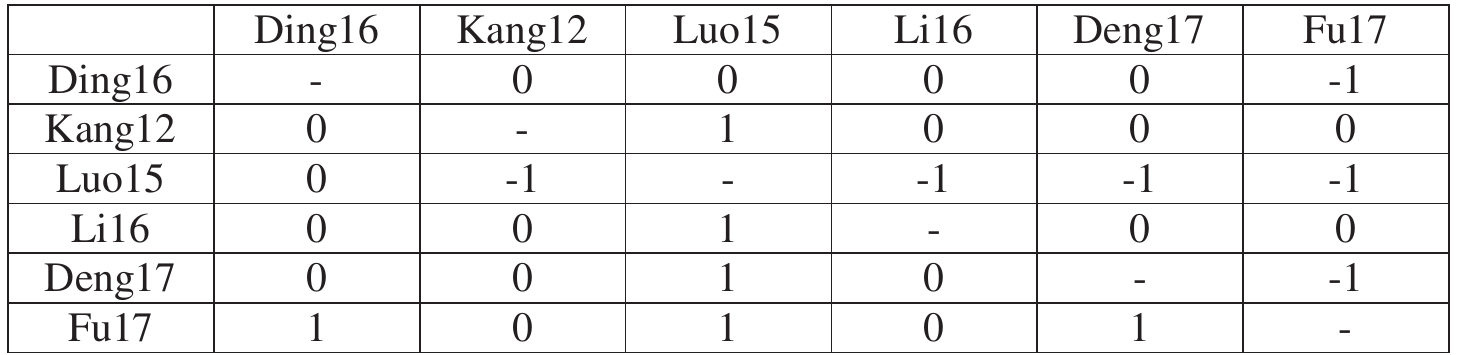}
  \label{table_ttest}
\end{table}

\begin{figure}[t]
  \centering
  \includegraphics[width=.85\linewidth]{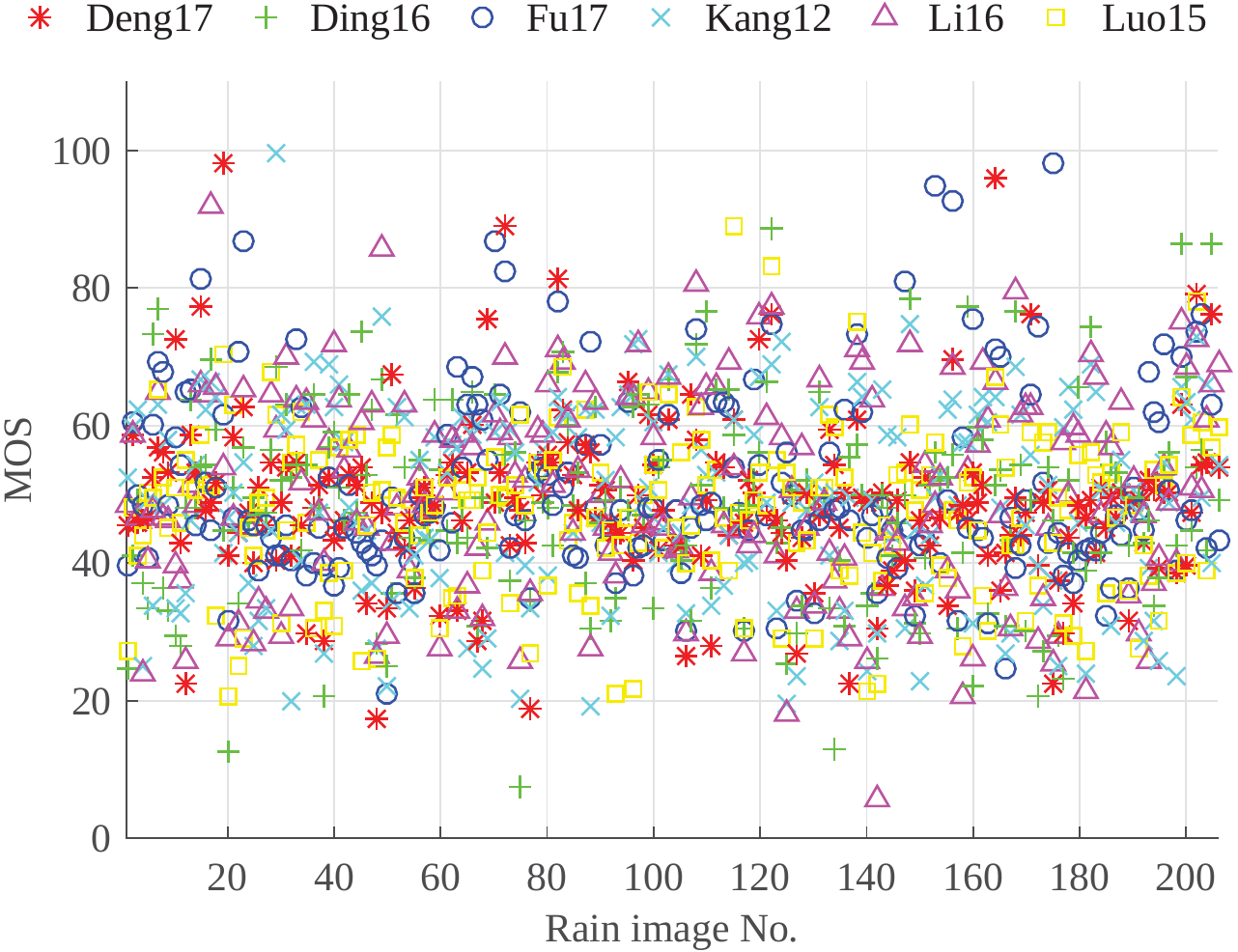}\\
  \caption{The MOS value of de-rained image versus its rain image number. The scatters with different shapes and colors indicate the de-rained images produced by different algorithms, which are labeled in the legend.}
  \label{fig-mos-vs-img}
\end{figure}

To quantitatively compare different de-raining algorithms, we also compute the mean and standard deviation for their MOS values. Each algorithm is associated to 206 MOS values, which are collected from its de-rained images. As shown in Fig. \ref{fig_mos_comparison}, we have two interesting findings. Firstly, in coping with the authentic rain images, the difference of overall performance is indistinguishable between existing de-raining algorithms, whose mean MOS values are very close to each other. To evaluate the statistical significance of this finding, we perform the onesided \textit{t-test} \cite{ttest} on the MOS values of each pair of de-raining algorithms. As shown in Table \ref{table_ttest}, the reported results confirm that most of de-raining algorithms are statistically equivalent to the others under 95\% confidence, which are denoted by `0'. The best performed method, i.e., Fu17, is only statistically superior to half of competitors including Ding16, Luo15 and Deng17. It shows a fact that there is no one de-raining algorithm possessing the absolute superiority with respect to the others in removing the realistic rain. Secondly, the de-raining performances of existing algorithms are unreliable, whose error bars are all quite large. It means that for any given de-raining algorithm, the perceptual qualities of its de-rained versions may change significantly from one image to another one. This finding could be verified from Fig. \ref{fig-mos-vs-img}, which plots the MOS value of each de-rained image versus its corresponding rain image number. It is seen that the scatters of each de-raining algorithm undulate largely across different rain images, which indicates that their performances are highly correlated with the rain types and image content.

\begin{figure*}[t]
  \centering
  \includegraphics[width=\linewidth]{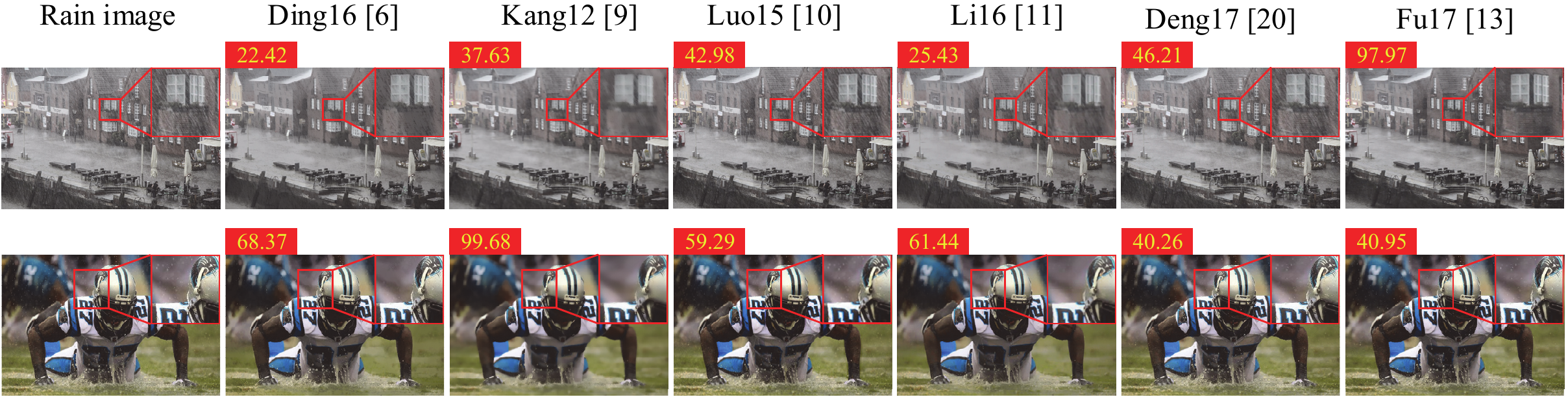}\\
  \caption{Illustration of the de-raining performance variation across different images. The MOS value of each de-rained image is labeled on its top-left corn.}
  \label{fig_performance_change}
\end{figure*}

To illustrate this problem, in Fig. \ref{fig_performance_change}, we show the de-raining results produced from two rain images. In the first row, the deep learning based method, i.e., Fu17 \cite{CNN_rain_removal}, performs well in removing the directional rain streaks and preserving the contours of window and balcony, whose MOS could reach 97.97. By contrast, the dictionary learning based method, such as, Kang12 \cite{removal_decomposition}, clearly over smoothes the original image structure, whose MOS is only 37.63. When we change the rain image to the second row, it is seen that Fu17 \cite{CNN_rain_removal} almost does nothing for the dot-like raindrops, whose MOS drops to 40.95. While, Kang12 \cite{removal_decomposition} could perfectly remove these small raindrops without obvious damage to the contour of the player, whose MOS rises to 99.68. This observation shows the pressing demand for an efficient NR-IQA model, which is crucial to select optimal de-raining algorithm in coping with different rain images.

\section{Objective model of DQA}

\begin{figure}[t]
  \centering
  \includegraphics[width=.9\linewidth]{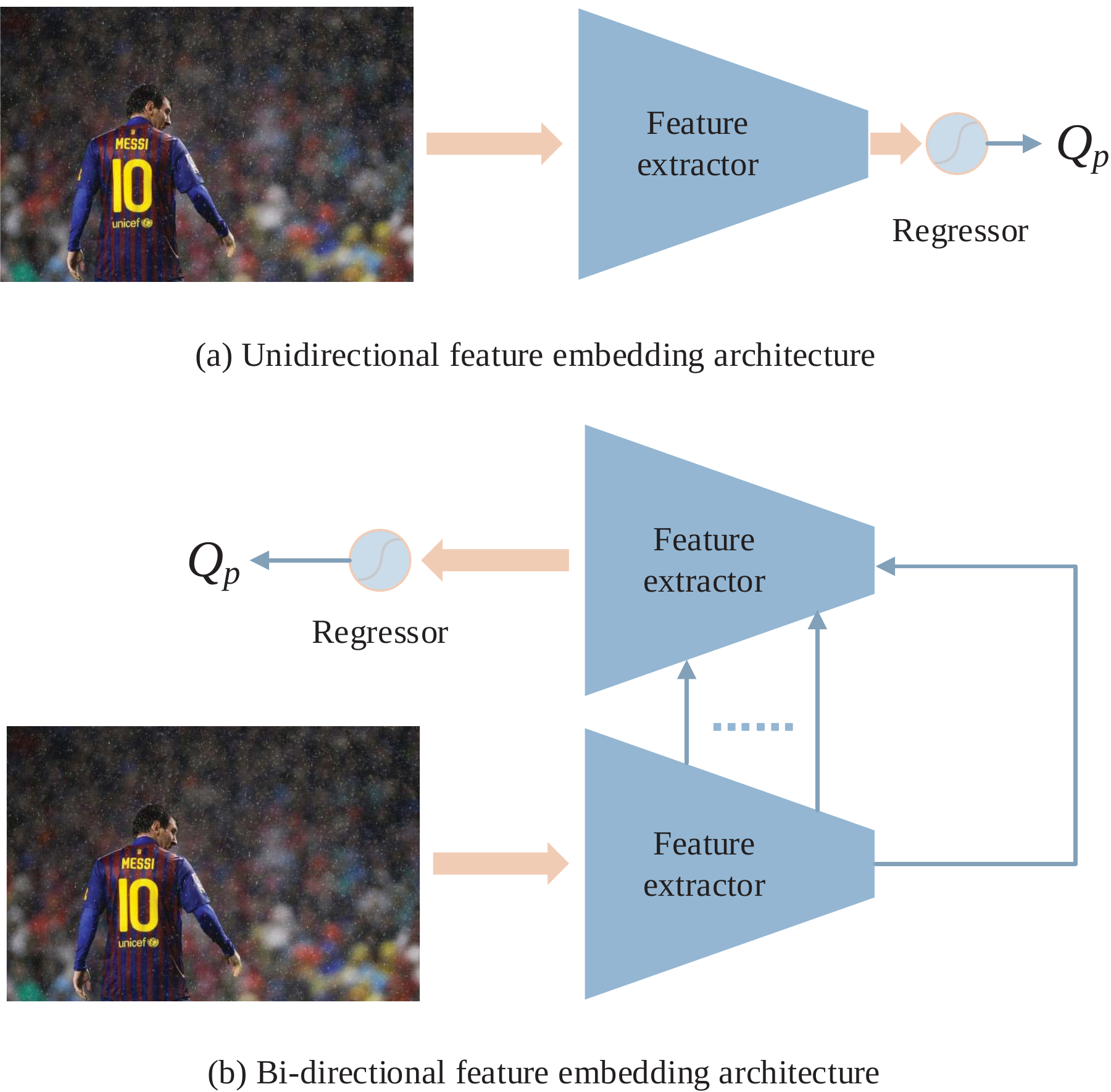}\\
  \caption{The comparison of different network architectures for NR-IQA.}
  \label{fig-architecture-comparison}
\end{figure}

\begin{figure*}[t]
  \centering
  \includegraphics[width=.75\linewidth]{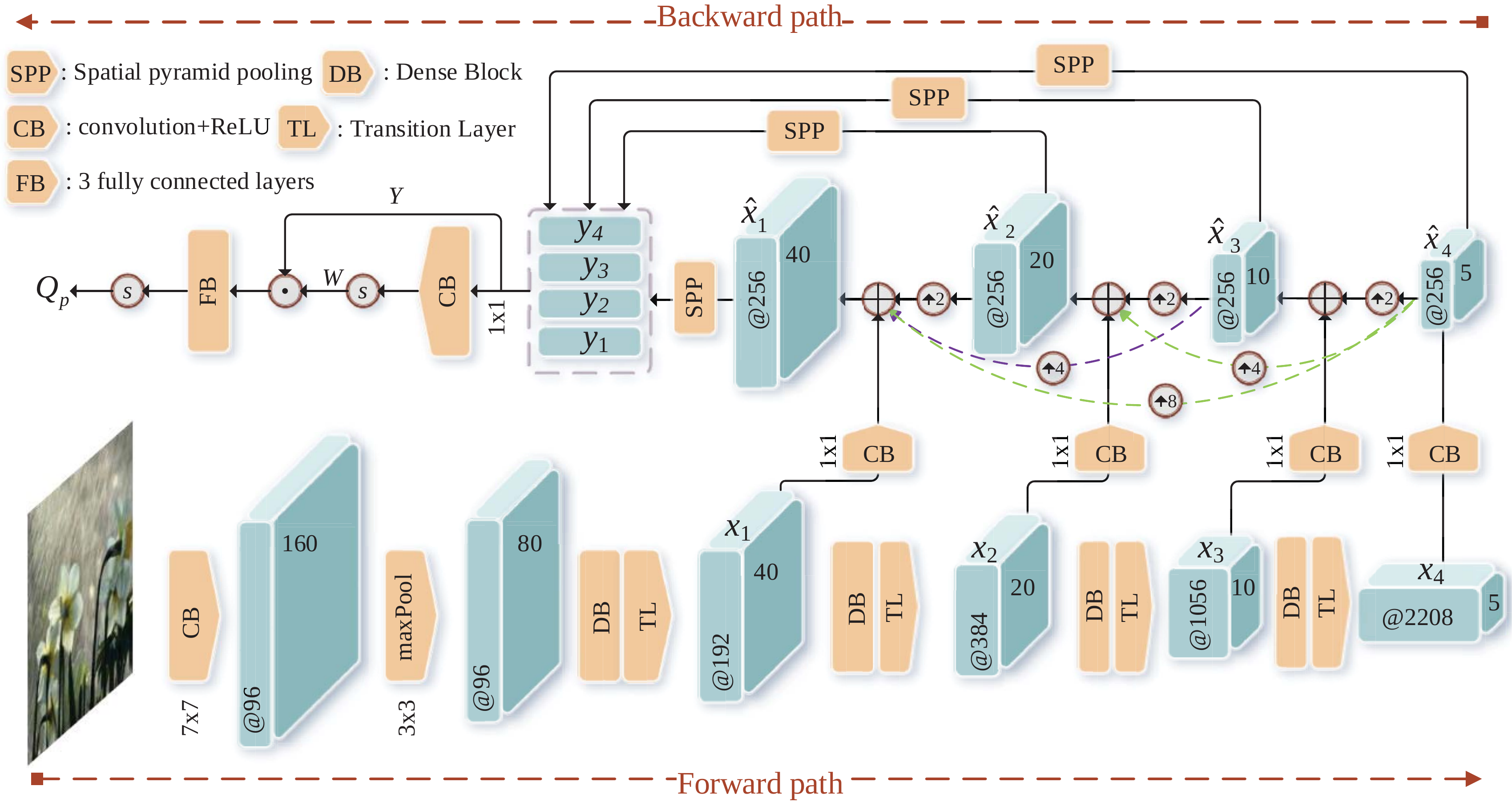}\\
  \caption{The detailed network structure of the proposed B-FEN. The purple and green dotted lines denote the features passed from $\hat{x}_3$ and $\hat{x}_4$, respectively.}
  \label{fig_framework}
\end{figure*}

After building the IVIPC-DQA database, we further develop an efficient objective model to predict the human perception towards the de-rained image. Recently, many deep learning based NR-IQA models \cite{MEON,DB-CNN,WaDIQaM} have explored various efficient network structures for evaluating the uniform distortions, which achieve state-of-the-art quality prediction accuracy via a common unidirectional feature embedding (UFE) architecture as shown in Fig. \ref{fig-architecture-comparison} (a). However,
unlike the typical distortions with distinct characteristic and uniform distribution (e.g., gaussian blur or white noise), the distortions of de-rained images are quite different across various de-raining algorithms and visual contents, which are hard to capture with specific global descriptor. Therefore, we propose to learn the quality-aware features and regressor by jointly considering the local and global information using a bi-directional feature embedding network (B-FEN) as shown in Fig. \ref{fig-architecture-comparison} (b). More specifically, the forward feature embedding aims to extract the perceptual quality related global information, which is similar to existing methods \cite{MEON,DB-CNN,WaDIQaM}. By contrast, the backward feature embedding attempts to incorporate the global information into multiple local features, which could be captured from the intermediate layers of a convolutional neural network (CNN) \cite{interpretation_CNN,dehazing-local,understanding_conv}.

For clarity, Fig. \ref{fig_framework} shows the detailed network structure of the proposed B-FEN. Our ``forward path'' subnetwork is composed of four cascaded Dense Blocks (DB) \cite{densenet}, which shares the same structure with DenseNet-161 except for the channel sizes. When a de-rained image goes deeper though our ``forward path'', the resolutions of the feature maps gradually decrease after a succession of pooling operations, which discard the semantic irrelevant information and squeeze more quality-aware global features into the top layer \cite{interpretation_CNN}. Specific to the DQA task, the global and uniform artifacts could be well captured from the top layer features, such as, the blurriness caused by the low-pass filter. However, the local information of image would be lost in this process, which plays a vital role in describing the non-uniform quality degradation across different regions.

Recently, many region- and pixel-level image representation methods \cite{ExFuse,understanding_conv,dehazing-local} have verified the efficiency of extracting local features from the intermediate-layer feature maps of a CNN. Inspired by these works, we further develop the ``backward path'' to unfold the way of incorporating the low-resolution global features into the high-resolution local features. In addition, since the importance of local and global features may vary across different image contents, we adopt a gated fusion module to adaptively determine the weights assigned to the multi-resolution feature maps, and merge them into a comprehensive feature vector to feed the quality regressor.

Let $X=\{x_i\}_{1\leq i\leq4}$ denote the ``forward path'' features outputted from four DB, where a larger $i$ denotes the deeper layer. Then, we reuse these feature maps in our ``backward path''. Let $\hat{X}=\{\hat{x}_i\}_{1\leq i\leq4}$ denote the features generated from the ``backward path''. Each element of $\hat{X}$ could be computed by integrating the feature maps from the current layer to the top layer, i.e.,
\begin{equation}
\hat{x}_i=
\begin{cases}
f_{1\times1}(x_i)+\sum_{j=i+1}^4\hat{x}_{j,\uparrow2^{j-i}} &1\leq i\leq3\\
f_{1\times1}(x_i) & \text{Otherwise}
\end{cases}
\end{equation}
where $f_{1\times1}(\cdot)$ denotes a $1\times1$ convolution and ReLU operation, and $\uparrow2^{j-i}$ denotes the upsampling operation with factor $2^{j-i}$, which is conducted by the transposed convolution. In this way, we sequentially embed the semantic-related information of top layer into the detail-related information of previous layers across different scales.

It is worth noting that the initial elements of $\hat{X}$ present different resolutions in simulating the perceptions of various receptive field sizes \cite{effectiveness_receptive_field}. Straightforwardly concatenating or merging these features would raise the bias towards high-resolution feature maps. To cope with this issue, as shown in Fig. \ref{fig_framework}, we first rescale all elements of $\hat{X}$ to the equal-length feature vectors via the spatial pyramid pooling (SPP) \cite{SPP}, which applies a regular 4$\times$4 and 2$\times$2 max-pooling window to each feature map and reshapes them to the feature vectors $\{y_i\}_{1\leq i\leq4}$. For clarity, we denote this operation by
\begin{equation}
y_i=\text{SPP}(\hat{x}_i)
\end{equation}
where the dimension of $y_i$ is 5120, i.e., (4$\times$4+2$\times$2)$\times$256.

Then, the weight of each $y_i$ could be computed as the nonlinear mapping of its response on the learnable $1\times1$ convolution, i.e.,
\begin{equation}
w_i=s[f_{1\times1}(y_i)]
\end{equation}
where $s(\cdot)$ is the sigmoid function and $w_i$ share the same dimension with $y_i$. For brevity, we stack the feature and weight vectors to the matrix form, i.e., $Y=[y_1;y_2;y_3;y_4]$ and $W=[w_1;w_2;w_3;w_4]$, which share the same dimension of 4$\times$5120. In the following, we assign the weights $W$ to $Y$ via an element-wise product operation, and generate the fused feature vector $z$ with a $1\times1$ convolution, i.e.,
\begin{equation}
z=f_{1\times1}(W\odot Y)
\end{equation}
where $\odot$ denotes the element-wise product operation.

Finally, the comprehensive feature vector $z$, which collects both the local and global quality-aware information, is fed to three cascaded fully connected (FC) layers and a sigmoid function to generate the predicted quality score $Q_p$. Let $Q_{gt}$ denote the ground-truth quality score. The learning objective of our B-FEN model is to minimize the $l_2$ loss between $Q_p$ and $Q_{gt}$, i.e.,
\begin{equation}
\mathcal{L}=\sum_{j=1}^N\frac{1}{N}\|Q_p(j)-Q_{gt}(j)\|_2^2
\end{equation}
where $N$ is the number of all training samples, $Q_p(j)$ and $Q_{gt}(j)$ denote the predicted and ground-truth quality scores of the $j$th de-rained image, respectively.

\section{Experiments}

To evaluate the performance of the proposed B-FEN model, we conduct the experiments on our IVIPC-DQA database. Due to the absence of specific quality metric for de-rained image, we compare the proposed B-FEN model with our previous B-GFN \cite{B-GFN} and some representative general-purpose image quality assessment models,  which include 10 opinion-aware (OA) metrics (i.e., BIQI \cite{BIQI}, BLIINDS II \cite{BLIINDS_II}, BRISQUE \cite{BRISQUE}, DIIVINE \cite{DIIVINE}, M3 \cite{M3}, NFERM \cite{NFERM}, TCLT \cite{TCLT}, MEON \cite{MEON}, DB-CNN \cite{DB-CNN} and WaDIQaM \cite{WaDIQaM}), and 4 opinion-unaware (OU) metrics (i.e., NIQE \cite{NIQE}, ILNIQE \cite{ILNIQE}, QAC \cite{QAC}, and LPSI \cite{LPSI}).  Meanwhile, two popular unidirectional feature embedding networks, i.e., DenseNet-161 \cite{densenet} and ResNet-152 \cite{resnet} are also involved in our comparison, which are categorized as OA metric in the following section.

\subsection{Implementation Details}

All OA metrics need training process to determine the parameters of quality assessment model. Following the setup of \cite{DB-CNN,MEON,WaDIQaM}, we randomly separate the IVIPC-DQA database into the non-overlapped training and testing sets, which include 80\% and 20\% images respectively.

For efficiently training our B-FEN model, the generic label preserving transformations including the random cropping and horizontal flipping \cite{data_augmentation} are used for augmenting the training data, where the cropped patch size is 320$\times$320. The four dense blocks are pre-trained on the ImageNet database \cite{ImageNet}, and the weights/biases of all the other convolutional layers are initialized by the recommendation of \cite{kaiming_uniform}. We employ the SGD optimizer \cite{SGD} for model learning and the mini-batch size is 16. The base learning rate is set to 0.01. In addition, the momentum and weight/bias decay parameters are set to 0.9 and 0, respectively.

\begin{table}[t]
\centering
  \caption{The evaluation results of all quality assessment models}
  \includegraphics[width=\linewidth]{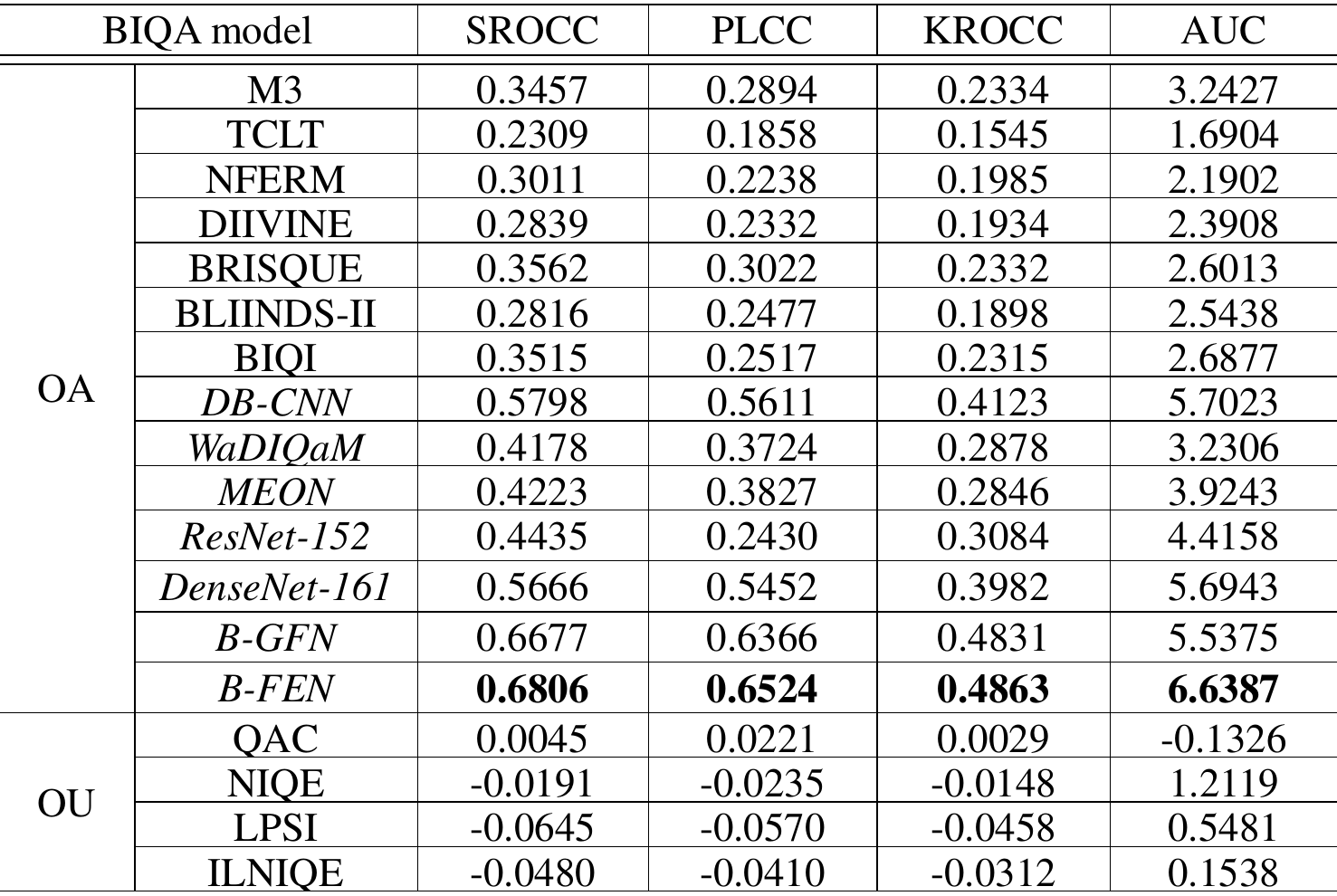}\\
  \label{table_consistency_result_all}
\end{table}

The DenseNet-161 \cite{densenet} and ResNet-152 \cite{resnet} models are pre-trained on the ImageNet database \cite{ImageNet} and then fine tuned on our IVIPC-DQA database. All the other OA metrics are directly re-trained on our IVIPC-DQA database, whose training settings follow the descriptions in their literatures \cite{BIQI,BLIINDS_II,BRISQUE,DIIVINE,M3,NFERM,TCLT,MEON,DB-CNN,WaDIQaM}. Since the OU metrics do not require quality labels to learn the parameters, we directly use the models released by the authors \cite{NIQE,ILNIQE,QAC,LPSI} to predict the image quality in the following experiments.

Similar to \cite{DB-CNN,MEON,WaDIQaM}, the random split is repeated 10 times and the median results of four popular indicators across all trials are reported for evaluating the DQA performance, which include the pearson's linear correlation coefficient (PLCC), spearman's rank correlation coefficient (SRCC), kendall's rank correlation coefficient (KRCC) and the perceptually weighted rank correlation (PWRC) \cite{PWRC}. It is noted that the PWRC indicator provides an overall performance measure, i.e., AUC, and a confidence-varying performance measure, i.e., \textit{SA-ST} curve.

\subsection{Consistency Evaluation}

In this section, we first compare the consistency between the subjective ratings and the predictions of different quality assessment algorithms towards the de-rained images. In Table \ref{table_consistency_result_all}, we report the overall prediction accuracies of all metrics, where the deep learning based methods are denoted by italics and the best results are highlighted by the boldface.

\begin{figure}[t]
  \centering
  \includegraphics[width=.8\linewidth]{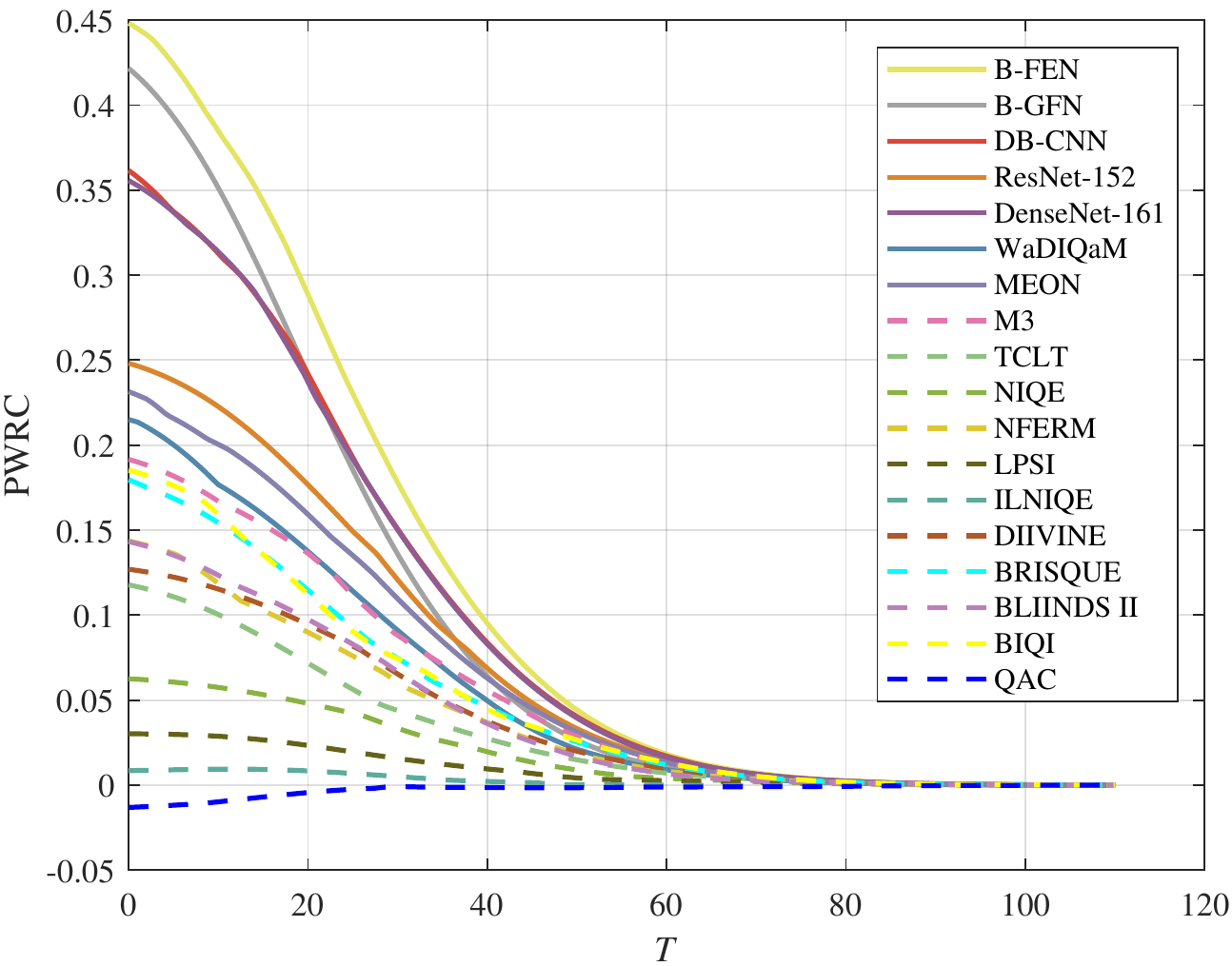}\\
  \caption{The SA-ST curves of different quality assessment models.}
  \label{fig-SAST-curve-consistency}
\end{figure}

It is seen that all of existing general-purpose NR-IQA models perform poorly in the DQA task. For the hand-crafted feature based models, their SRCC are all smaller than 0.4. Meanwhile, limited by the image prior learned from rainless scenes and inflexible parameter settings, the OU metrics produce much worse performance, whose SRCC are all even close to or smaller than 0. It shows that the DQA task is more challenging than the traditional uniform distortion evaluation, where the OU metrics could achieve comparable or even better performance than the OA metrics \cite{NIQE,ILNIQE,QAC,LPSI}.

\begin{figure*}[t]
  \centering
  \includegraphics[width=\linewidth]{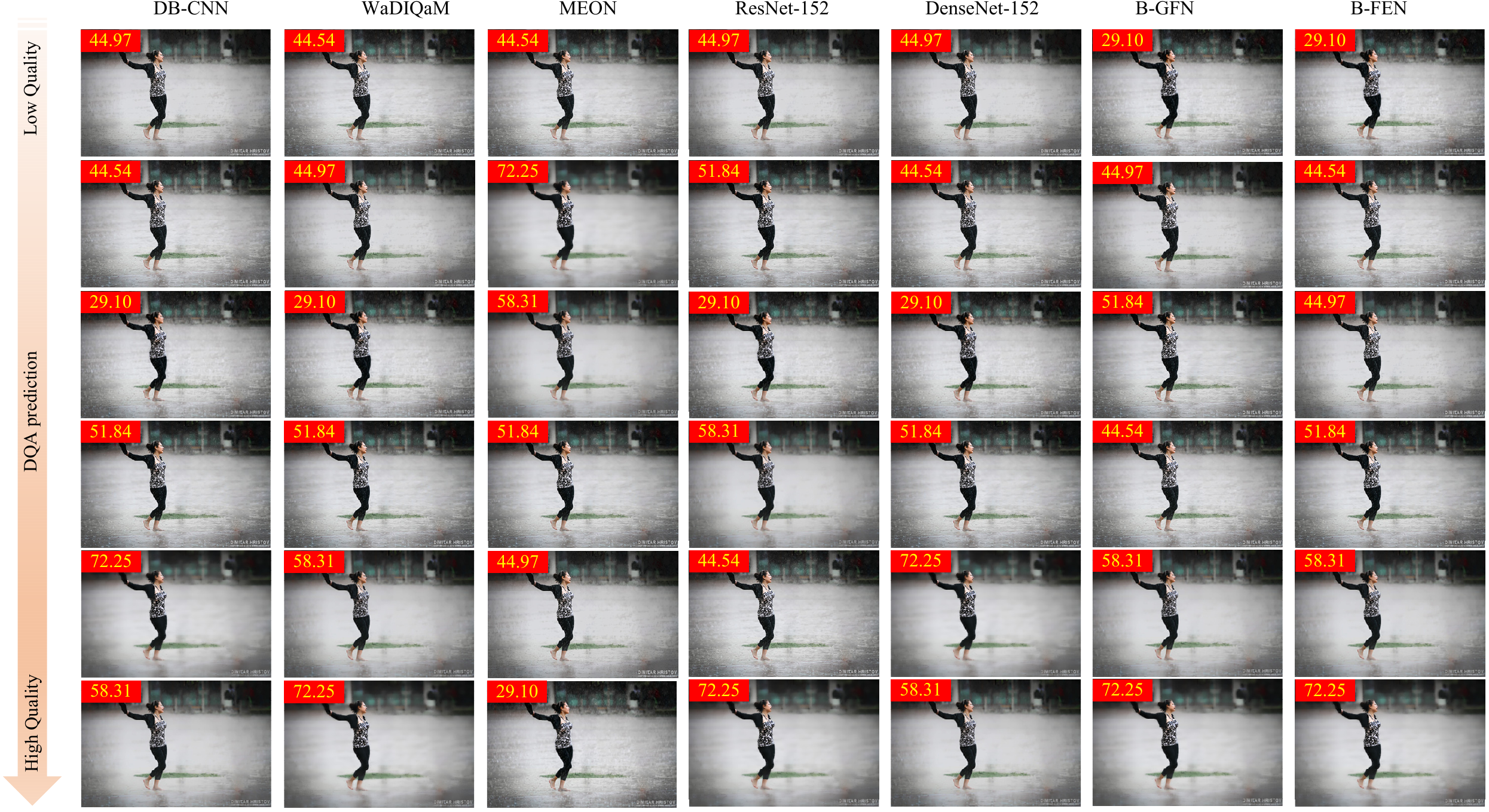}\\
  \caption{Illustration of ranking results from different deep image quality assessment models. In each column, six de-rained versions of an image are ranked in descending order from top to down by each DQA model. The MOS value of each image is labeled in its top left corner.}
  \label{fig-rank-illustration}
\end{figure*}

Due to the great capability of joint learning discriminative features and regressors, all deep learning based OA metrics report much better results, whose SRCC are larger than 0.4. More specifically, the MEON, WaDIQaM and ResNet-152 directly pass the feature maps from the shallow layer to the deep layer, whose performance improvements are still moderate. The DenseNet-161 and DB-CNN further enhance the quality-aware global representation by feature reuse and fusion. Due to this superiority, the SRCC values of ResNet-152 and DB-CNN raise up to 0.57, which outperform the MEON and WaDIQaM. It is worth noting that these representative deep learning based NR-IQA models (such as, MEON, DB-CNN and WaDIQaM) and the popular CNNs  (such as, DenseNet-161 and ResNet-152) all employ an unidirectional feature embedding architecture, whose feature map size gradually decreases from the shallow layer to the deep layer and the local information are erased after successive pooling operations. By contrast, we incorporate the local details into the global features via our unique bi-directional feature embedding network, which is quite beneficial for describing the non-uniform distortions in the de-rained image. Finally, our B-GFN and B-FEN models produce much better quality prediction results in terms of all indicators, whose SRCCs exceed 0.6 and approach 0.7. In addition, since the B-FEN enhances the feature reuse in the `backward path', we report better performance than our previous B-FGN model.

Fig. \ref{fig-SAST-curve-consistency} further plots the \textit{SA-ST} curves of different quality assessment models, where the deep and handcrafted IQA models are labeled by the solid and dotted lines respectively. It is seen that our B-FEN significantly outperforms existing general-purpose NR-IQA models and our previous B-GFN across a wide range of confidence interval. That is, we perform better in correctly ranking the high quality image pairs no matter their perceptual difference is small or large \cite{PWRC}. This is important in recommending perceptually preferred de-raining results in various real-world applications. For clarity, an illustration of ranking results from different deep image quality assessment models is given in Fig. \ref{fig-rank-illustration}, where our B-FEN model produces the same quality rank with respect to the human ratings and all the other models show several rank errors.

\subsection{Rain-remover Independency}

Besides the consistency investigation for the proposed objective model, we also conduct the leave-one-out cross validation \cite{cross-validation} to verify that the accuracy of our B-FEN is not dependent on any specific de-raining algorithm. More specifically, for each de-raining algorithm, we take its 206 de-rained images as the test set and the rest images produced by all the other de-raining algorithms are used for training our B-FEN model. We repeat this trial 6 times until all de-raining algorithms are separately tested in our experiment. Let $C_i$ denote the overall performance of the $i$th quality assessment model, and $A_{i,j}$ denote the accuracy of the $i$th quality assessment model towards the $j$th de-raining algorithm. Following the criteria of \cite{cross-validation}, we represent the $C_i$ by
\begin{equation}
C_i = \frac{1}{6}\sum_{j=1}^6A_{i,j}
\end{equation}
where this overall performance is computed across all indicators, i.e., SRCC/PLCC/KRCC/PWRC.

\begin{table}[t]
\centering
  \caption{The evaluation results of all quality assessment models}
  \includegraphics[width=\linewidth]{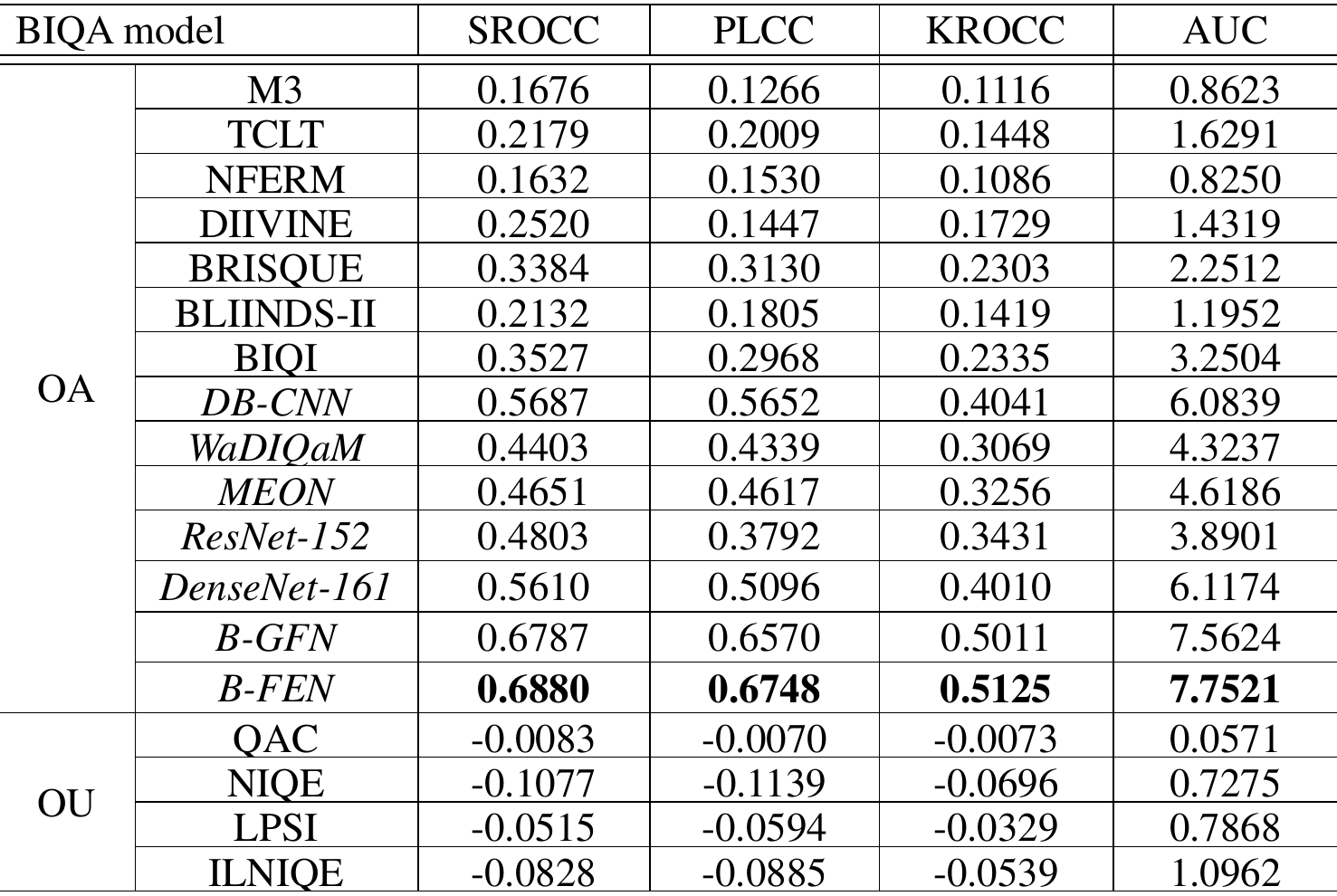}\\
  \label{table_consistency_result_ind}
\end{table}

\begin{figure}[t]
  \centering
  \includegraphics[width=.8\linewidth]{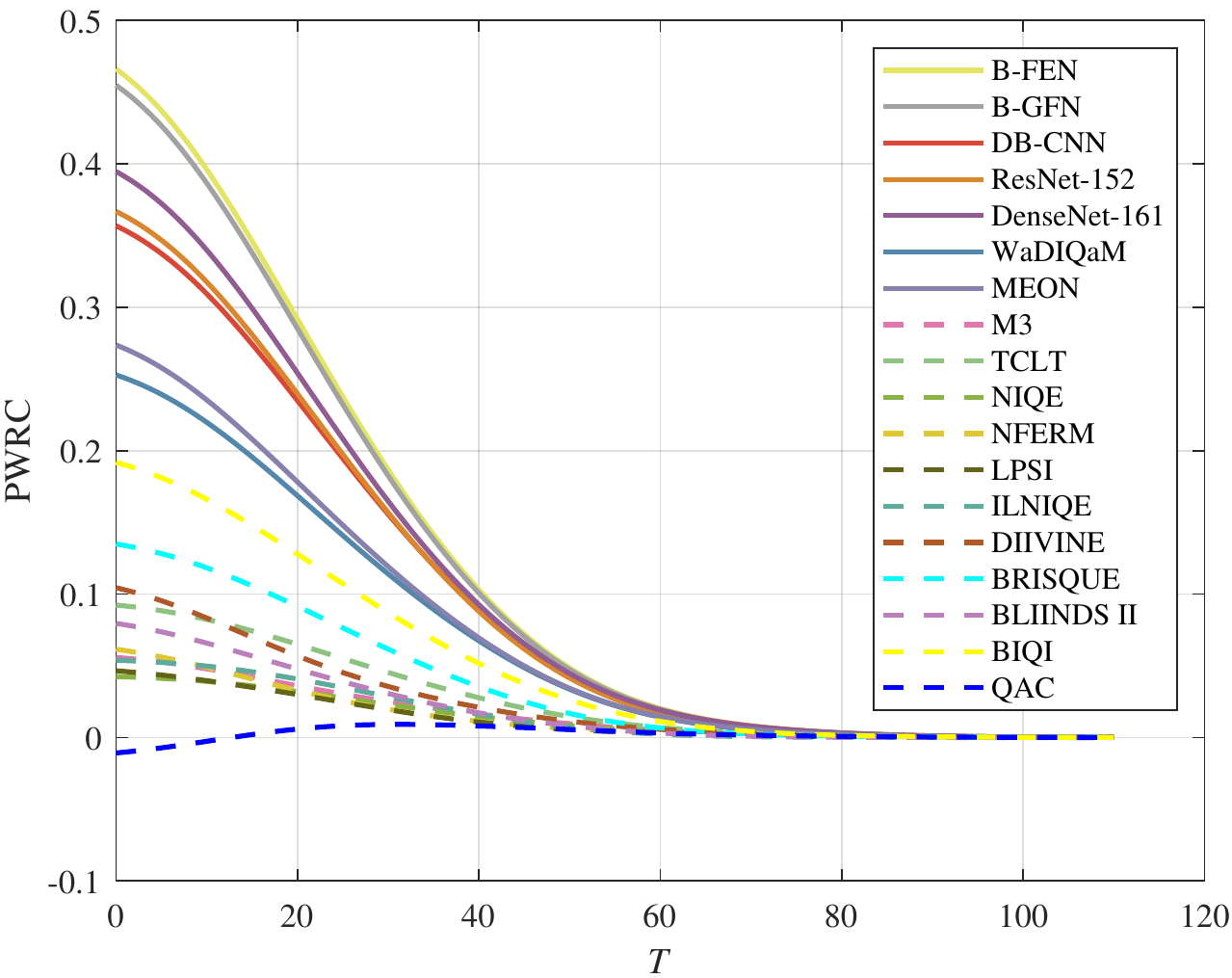}\\
  \caption{The SA-ST curves for rain-remover independency investigation.}
  \label{fig-SAST-curve-ind}
\end{figure}

Table \ref{table_consistency_result_ind} shows the independency investigation results for all quality assessment models. It is seen that the deep learning based methods still perform better than the handcrafted models, and the OA metrics significantly outperform the OU metrics. This demonstrates that a powerful learning capability is necessary for bridging the gap from the uniform distortion measure to the nonuniform distortion measure.

In addition, we can find that the straight through UFE networks, i.e., MEON, WaDIQaM and ResNet-152, are still inferior to the feature fusion based UFE networks, such as, DenseNet-161 and DB-CNN. Meanwhile, the proposed B-FEN also performs best in the independency investigation. It confirms that a more comprehensive local and global quality-aware feature representation is the key for DQA no matter which deraining algorithm is applied to the rainy image.

Similar to Section IV-B, we also show the \textit{SA-ST} curves of different quality assessment models in this rain-remover independency investigation. As shown in Fig. \ref{fig-SAST-curve-ind}, our B-FEN still outperforms all the other NR-IQA models across different confidence intervals. It demonstrates that the proposed DQA model offers better de-rained recommendations to the users no matter which de-raining algorithm is used in current application.

\subsection{Complexity Analysis}

\begin{table*}[t]
\centering
  \caption{Complexity analysis for deep quality assessment models. `M' and `B' represent the units of million and billion, respectively.}
  \includegraphics[width=.8\linewidth]{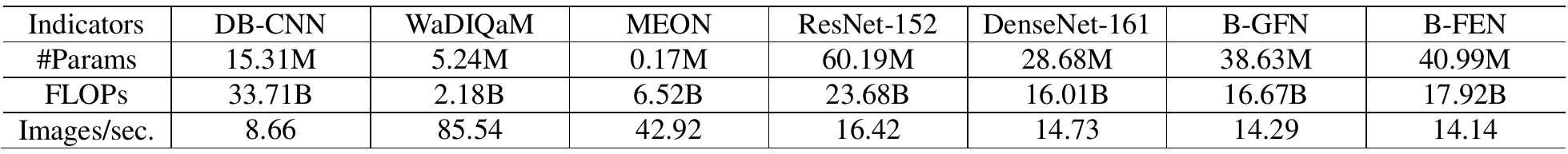}\\
  \label{table_complexity}
\end{table*}

Besides the evaluation accuracy, we further compare the complexities of different deep quality assessment methods by the number of parameters (\#Params.), floating-point operations per second (FLOPs) \cite{complexity}, and actual running speed (Images/sec.). We implement the proposed B-FEN method with the PyTorch
library, and perform the experiments in a workstation
with Intel Xeon E5-2660 CPU and NVIDIA TITAN X GPU.

\begin{figure*}[t]
\centering
  \subfigure[SROCC versus \#Params.]{
  \includegraphics[width=.32\linewidth]{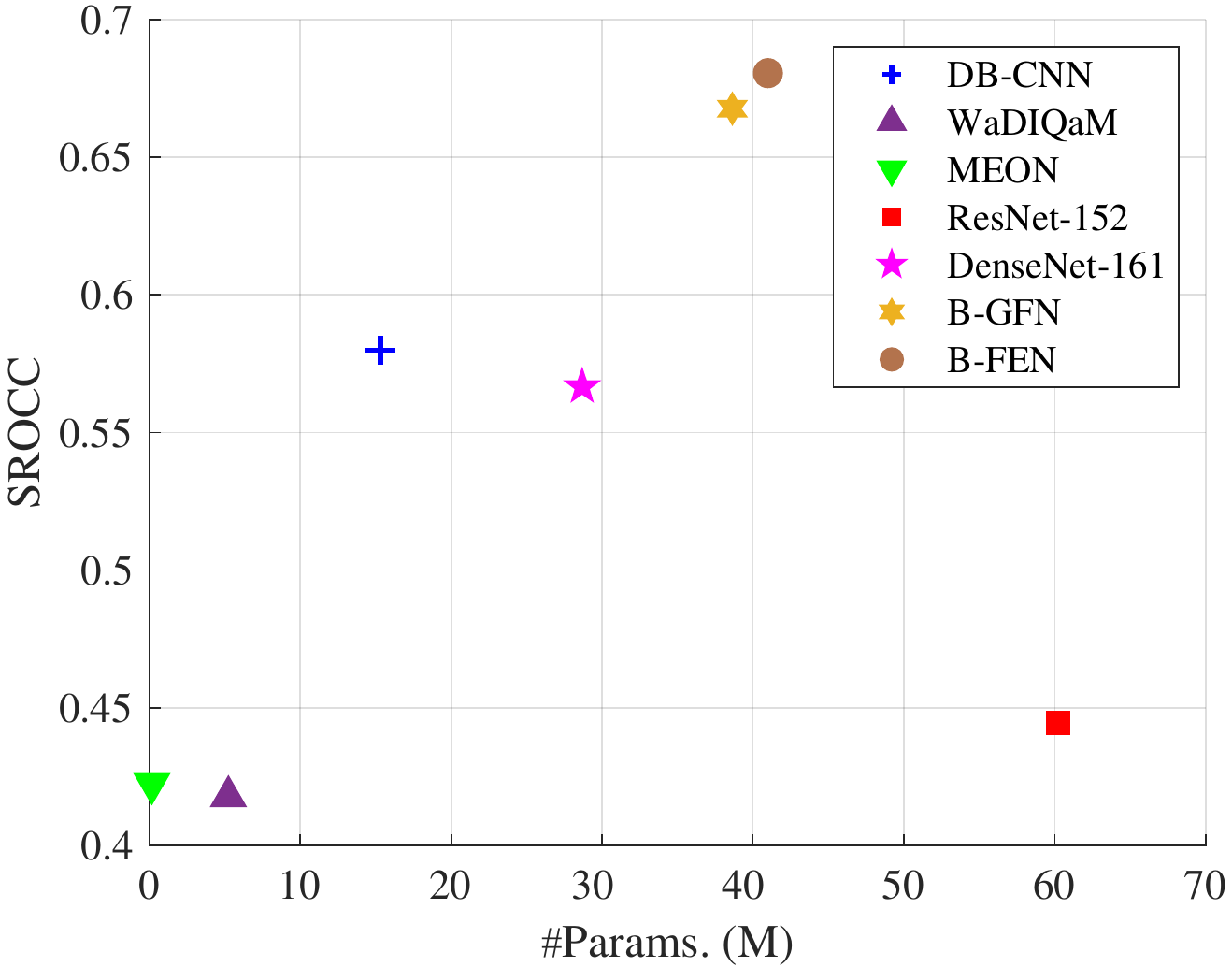}}
  \subfigure[SROCC versus FLOPs.]{
  \includegraphics[width=.32\linewidth]{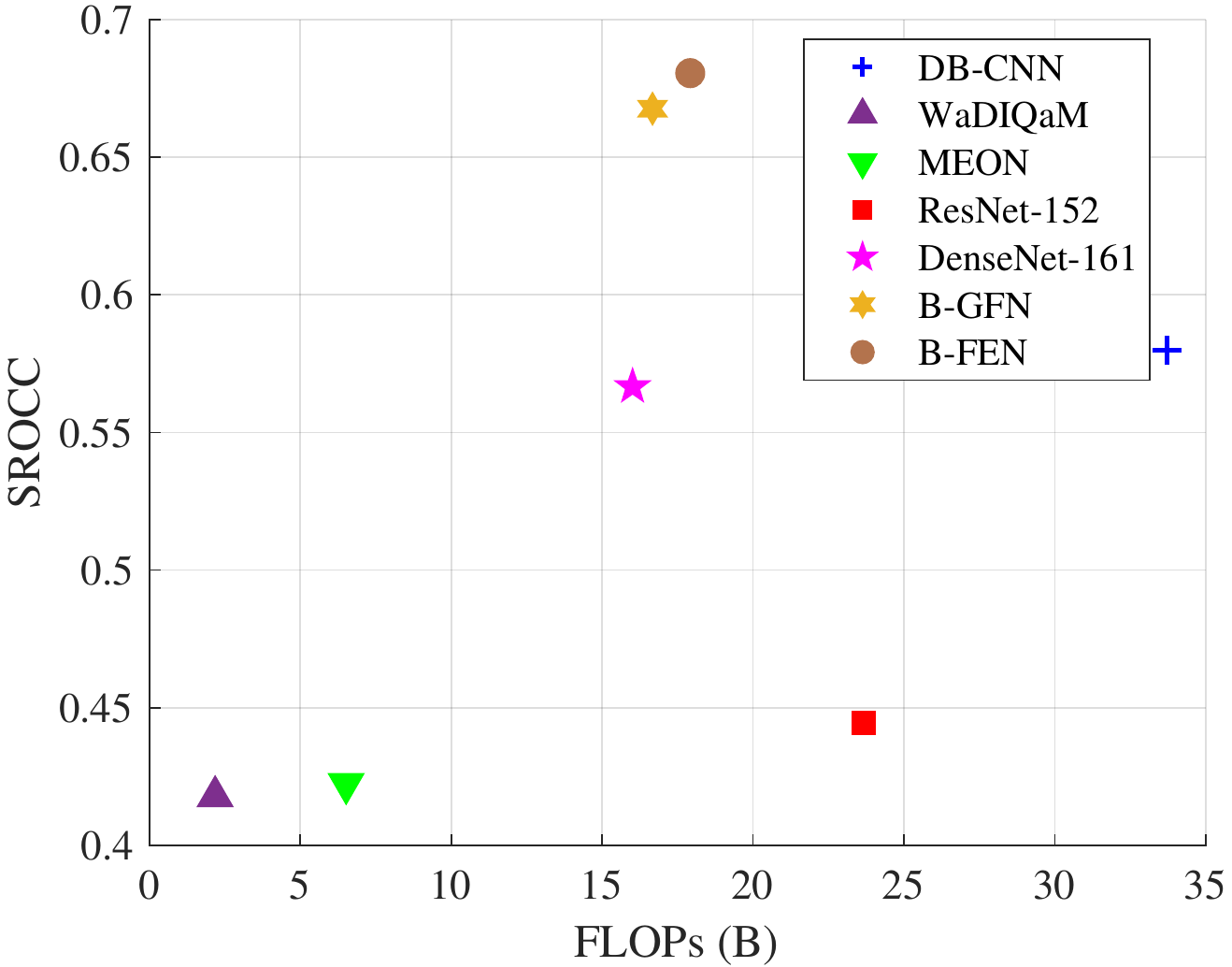}}
  \subfigure[SROCC versus Images/sec.]{
  \includegraphics[width=.32\linewidth]{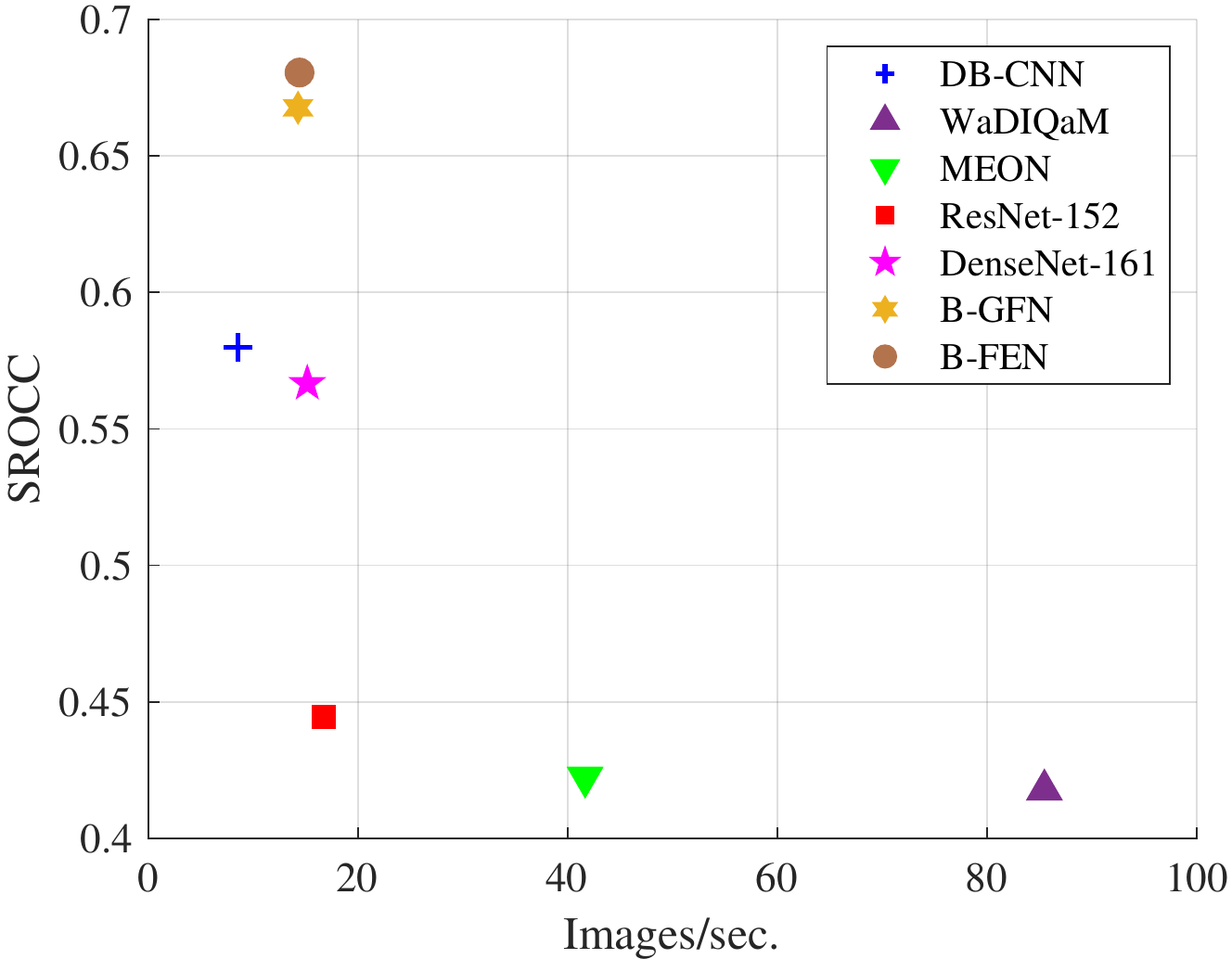}}
  \caption{Performance and complexity analysis for deep quality assessment models. `M' and `B' represent the units of million and billion, respectively.}
  \label{fig_complexity}
\end{figure*}

Table \ref{table_complexity} shows the statistical results of all deep IQA models. It is seen that the MEON and WaDIQaM show the lowest complexities, which are developed with shallow network architectures and report the smallest \#Params. and FLOPs. In addition, the complexity of the proposed B-FEN is moderate, whose \#Params. is smaller than ResNet-152 and the FLOPs is lower than both ResNet-152 and DB-CNN. This benefits from the application of multiple 1$\times$1 convolutions and limited channels in our backward path, which only slightly increase the parameter size and computation cost in comparison with the DenseNet-161. Meanwhile, the proposed B-FEN adds more feature reuse in the `backward path', which leads to a litter higher complexity than our previous B-GFN. The DB-CNN uses fewer convolution layers and presents smaller \#Params. than the proposed model. However, since multiple high dimensional features are employed in the fully connected layers, the FLOPs of DB-CNN is significantly higher than our B-FEN.
Finally, the running speed of our B-FEN model could reach 14.14 Images/sec., which is close to DenseNet-161 and much faster than DB-CNN.

Fig. \ref{fig_complexity} further investigates the relationship between the performance and complexity for the deep IQA models, where the SROCC is used as the performance indicator and \#Params., FLOPs, Images/sec. are used for complexity measurement. We can find that the proposed B-FEN performs well in balancing the evaluation accuracy and complexity. More specifically, we achieve the highest SROCC with moderate memory and computation costs.

\section{Conclusion}

Single image rain removal has received extensive attentions recently. However, there is very few work dedicated to the quality assessment of de-rained images. In this paper, we first build a new database to collect the human rated scores for the de-rained versions of various authentic rain images. Then, a bi-directional feature embedding network (B-FEN) is proposed to predict the human perception toward different de-rained images. Experimental results show that the de-raining quality assessment task is quite challenging and all of existing general purpose BIQA models fail to accurately predict the perceptual de-raining quality. By means of the enriched global and local feature representation, our proposed B-FEN produces very promising DQA result, which significantly outperforms many representative BIQA models and the state-of-the-art deep neural networks. Our new database and B-FEN metric are helpful for evaluating and developing the perceptually preferred de-raining algorithms in the authentic rain scenes.

%

\ifCLASSOPTIONcaptionsoff
  \newpage
\fi



\bibliographystyle{IEEEtran}
\bibliography{refs}

\begin{thebibliography}{10}
\providecommand{\url}[1]{#1}
\csname url@samestyle\endcsname
\providecommand{\newblock}{\relax}
\providecommand{\bibinfo}[2]{#2}
\providecommand{\BIBentrySTDinterwordspacing}{\spaceskip=0pt\relax}
\providecommand{\BIBentryALTinterwordstretchfactor}{4}
\providecommand{\BIBentryALTinterwordspacing}{\spaceskip=\fontdimen2\font plus
\BIBentryALTinterwordstretchfactor\fontdimen3\font minus
  \fontdimen4\font\relax}
\providecommand{\BIBforeignlanguage}[2]{{%
\expandafter\ifx\csname l@#1\endcsname\relax
\typeout{** WARNING: IEEEtran.bst: No hyphenation pattern has been}%
\typeout{** loaded for the language `#1'. Using the pattern for}%
\typeout{** the default language instead.}%
\else
\language=\csname l@#1\endcsname
\fi
#2}}
\providecommand{\BIBdecl}{\relax}
\BIBdecl

\bibitem{depth_deraining}
D.~{Chen}, C.~{Chen}, and L.~{Kang}, ``Visual depth guided color image rain
  streaks removal using sparse coding,'' \emph{IEEE Transactions on Circuits
  and Systems for Video Technology}, vol.~24, no.~8, pp. 1430--1455, Aug 2014.

\bibitem{deraining_GAN}
H.~{Zhang}, V.~{Sindagi}, and V.~M. {Patel}, ``Image de-raining using a
  conditional generative adversarial network,'' \emph{IEEE Transactions on
  Circuits and Systems for Video Technology}, pp. 1--1, 2019, in Press.

\bibitem{rain_analysis}
P.~C. Barnum, S.~Narasimhan, and T.~Kanade, ``Analysis of rain and snow in
  frequency space,'' \emph{International Journal of Computer Vision}, vol.~86,
  no.~2, p. 256, Jan 2009.

\bibitem{image_restoration}
A.~K. Katsaggelos, \emph{Digital Image Restoration}.\hskip 1em plus 0.5em minus
  0.4em\relax Springer Publishing Company, Incorporated, 2012.

\bibitem{nonlocal_sparse_restoration}
H.~{Liu}, R.~{Xiong}, X.~{Zhang}, Y.~{Zhang}, S.~{Ma}, and W.~{Gao}, ``Nonlocal
  gradient sparsity regularization for image restoration,'' \emph{IEEE
  Transactions on Circuits and Systems for Video Technology}, vol.~27, no.~9,
  pp. 1909--1921, Sep. 2017.

\bibitem{guided_filter}
K.~He, J.~Sun, and X.~Tang, ``Guided image filtering,'' \emph{IEEE TPAMI},
  vol.~35, no.~6, pp. 1397--1409, June 2013.

\bibitem{guided_filter_removal}
X.~Ding, L.~Chen, X.~Zheng, Y.~Huang, and D.~Zeng, ``Single image rain and snow
  removal via guided l0 smoothing filter,'' \emph{Multimedia Tools and
  Applications}, vol.~75, no.~5, pp. 2697--2712, Mar 2016.

\bibitem{multi-filter}
X.~Zheng, Y.~Liao, W.~Guo, X.~Fu, and X.~Ding, ``Single-image-based rain and
  snow removal using multi-guided filter,'' in \emph{ICONIP}, 2013, pp.
  258--265.

\bibitem{low_rank_model}
Y.~L. Chen and C.~T. Hsu, ``A generalized low-rank appearance model for
  spatio-temporally correlated rain streaks,'' in \emph{IEEE International
  Conference on Computer Vision}, Dec 2013, pp. 1968--1975.

\bibitem{removal_decomposition}
L.~W. Kang, C.~W. Lin, and Y.~H. Fu, ``Automatic single-image-based rain
  streaks removal via image decomposition,'' \emph{IEEE Transactions on Image
  Processing}, vol.~21, no.~4, pp. 1742--1755, April 2012.

\bibitem{DSP_rain_removal}
Y.~Luo, Y.~Xu, and H.~Ji, ``Removing rain from a single image via
  discriminative sparse coding,'' in \emph{IEEE International Conference on
  Computer Vision}, Dec 2015, pp. 3397--3405.

\bibitem{layer_prior_removal}
Y.~Li, R.~T. Tan, X.~Guo, J.~Lu, and M.~S. Brown, ``Rain streak removal using
  layer priors,'' in \emph{IEEE Conference on Computer Vision and Pattern
  Recognition}, June 2016, pp. 2736--2744.

\bibitem{joint_detection_removal}
\BIBentryALTinterwordspacing
W.~Yang, R.~T. Tan, J.~Feng, J.~Liu, Z.~Guo, and S.~Yan, ``Joint rain detection
  and removal via iterative region dependent multi-task learning,''
  \emph{CoRR}, vol. abs/1609.07769, 2016. [Online]. Available:
  \url{http://arxiv.org/abs/1609.07769}
\BIBentrySTDinterwordspacing

\bibitem{CNN_rain_removal}
X.~Fu, J.~Huang, X.~Ding, Y.~Liao, and J.~Paisley, ``Clearing the skies: A deep
  network architecture for single-image rain removal,'' \emph{IEEE Transactions
  on Image Processing}, vol.~26, no.~6, pp. 2944--2956, June 2017.

\bibitem{SSIM}
Z.~Wang, A.~C. Bovik, H.~R. Sheikh, and E.~P. Simoncelli, ``Image quality
  assessment: From error visibility to structural similarity,'' \emph{IEEE
  Transactions on Image Processing}, vol.~13, no.~4, pp. 600--612, Apr. 2004.

\bibitem{B-GFN}
Q.~{Wu}, L.~{Wang}, K.~N. {Ngan}, H.~{Li}, and F.~{Meng}, ``Beyond synthetic
  data: A blind deraining quality assessment metric towards authentic rain
  image,'' in \emph{IEEE International Conference on Image Processing}, Sep.
  2019, pp. 2364--2368.

\bibitem{LOCRUE}
Q.~{Wu}, H.~{Li}, K.~N. {Ngan}, and K.~{Ma}, ``Blind image quality assessment
  using local consistency aware retriever and uncertainty aware evaluator,''
  \emph{IEEE Transactions on Circuits and Systems for Video Technology},
  vol.~28, no.~9, pp. 2078--2089, Sep. 2018.

\bibitem{BIQI}
A.~K. Moorthy and A.~C. Bovik, ``A two-step framework for constructing blind
  image quality indices,'' \emph{IEEE Signal Processing Letter}, vol.~17,
  no.~5, pp. 513--516, May 2010.

\bibitem{TCLT}
Q.~Wu, H.~Li, F.~Meng, K.~N. Ngan, B.~Luo, C.~Huang, and B.~Zeng, ``Blind image
  quality assessment based on multichannel feature fusion and label transfer,''
  \emph{IEEE Transactions on Circuits and Systems for Video Technology},
  vol.~26, no.~3, pp. 425--440, March 2016.

\bibitem{R3}
Q.~Wu, H.~Li, Z.~Wang, F.~Meng, B.~Luo, W.~Li, and K.~N. Ngan, ``Blind image
  quality assessment based on rank-order regularized regression,'' \emph{IEEE
  Transactions on Multimedia}, vol.~19, no.~11, pp. 2490--2504, Nov 2017.

\bibitem{DIIVINE}
A.~K. Moorthy and A.~C. Bovik, ``Blind image quality assessment: From natural
  scene statistics to perceptual quality,'' \emph{IEEE Transactions on Image
  Processing}, vol.~20, no.~12, pp. 3350--3364, Dec 2011.

\bibitem{BRISQUE}
A.~Mittal, A.~K. Moorthy, and A.~C. Bovik, ``No-reference image quality
  assessment in the spatial domain,'' \emph{IEEE Transactions on Image
  Processing}, vol.~21, no.~12, pp. 4695--4708, Dec 2012.

\bibitem{MSGF-PR}
Q.~Wu, H.~Li, F.~Meng, K.~N. Ngan, and S.~Zhu, ``No reference image quality
  assessment metric via multi-domain structural information and piecewise
  regression,'' \emph{J. Vis. Commun Image R.}, vol.~32, no. Supplement C, pp.
  205 -- 216, 2015.

\bibitem{UGSM}
L.~Deng, T.~Huang, X.~Zhao, and T.~Jiang, ``A directional global sparse model
  for single image rain removal,'' \emph{Applied Mathematical Modelling},
  vol.~59, pp. 662--679, 2018.

\bibitem{BT500}
ITU-R, ``Recommendation bt.500-13: Methodology for subjective assessment of the
  quality of television pictures,'' [Online] Available:
  \url{https://www.itu.int/rec/R-REC-BT.500-13-201201-I/en}, 2012.

\bibitem{P910}
ITU-T, ``Recommendation p.910: Subjective video quality assessment methods for
  multimedia applications,'' [Online] Available:
  \url{https://www.itu.int/rec/T-REC-P.910-200804-I/en}, 2008.

\bibitem{score_range}
K.~Seshadrinathan, R.~Soundararajan, A.~C. Bovik, and L.~K. Cormack, ``Study of
  subjective and objective quality assessment of video,'' \emph{IEEE
  Transactions on Image Processing}, vol.~19, no.~6, pp. 1427--1441, Jun. 2010.

\bibitem{retarget}
L.~{Ma}, W.~{Lin}, C.~{Deng}, and K.~N. {Ngan}, ``Image retargeting quality
  assessment: A study of subjective scores and objective metrics,'' \emph{IEEE
  Journal of Selected Topics in Signal Processing}, vol.~6, no.~6, pp.
  626--639, Oct 2012.

\bibitem{wireless_video}
A.~K. {Moorthy}, K.~{Seshadrinathan}, R.~{Soundararajan}, and A.~C. {Bovik},
  ``Wireless video quality assessment: A study of subjective scores and
  objective algorithms,'' \emph{IEEE Transactions on Circuits and Systems for
  Video Technology}, vol.~20, no.~4, pp. 587--599, April 2010.

\bibitem{ttest}
D.~J. Sheskin, \emph{Handbook of parametric and nonparametric statistical
  procedures}.\hskip 1em plus 0.5em minus 0.4em\relax CRC Press, 2003.

\bibitem{MEON}
K.~{Ma}, W.~{Liu}, K.~{Zhang}, Z.~{Duanmu}, Z.~{Wang}, and W.~{Zuo},
  ``End-to-end blind image quality assessment using deep neural networks,''
  \emph{IEEE Transactions on Image Processing}, vol.~27, no.~3, pp. 1202--1213,
  March 2018.

\bibitem{DB-CNN}
W.~{Zhang}, K.~{Ma}, J.~{Yan}, D.~{Deng}, and Z.~{Wang}, ``Blind image quality
  assessment using a deep bilinear convolutional neural network,'' \emph{IEEE
  Transactions on Circuits and Systems for Video Technology}, pp. 1--1, 2018.

\bibitem{WaDIQaM}
S.~{Bosse}, D.~{Maniry}, K.~{M¨¹ller}, T.~{Wiegand}, and W.~{Samek}, ``Deep
  neural networks for no-reference and full-reference image quality
  assessment,'' \emph{IEEE Transactions on Image Processing}, vol.~27, no.~1,
  pp. 206--219, Jan 2018.

\bibitem{interpretation_CNN}
\BIBentryALTinterwordspacing
N.~Akhtar and U.~Ragavendran, ``Interpretation of intelligence in cnn-pooling
  processes: a methodological survey,'' \emph{Neural Computing and
  Applications}, Jul 2019. [Online]. Available:
  \url{https://doi.org/10.1007/s00521-019-04296-5}
\BIBentrySTDinterwordspacing

\bibitem{dehazing-local}
W.~Ren, S.~Liu, H.~Zhang, J.~Pan, X.~Cao, and M.-H. Yang, ``Single image
  dehazing via multi-scale convolutional neural networks,'' in \emph{European
  conference on computer vision}.\hskip 1em plus 0.5em minus 0.4em\relax
  Springer, 2016, pp. 154--169.

\bibitem{understanding_conv}
P.~{Wang}, P.~{Chen}, Y.~{Yuan}, D.~{Liu}, Z.~{Huang}, X.~{Hou}, and
  G.~{Cottrell}, ``Understanding convolution for semantic segmentation,'' in
  \emph{IEEE Winter Conference on Applications of Computer Vision}, March 2018,
  pp. 1451--1460.

\bibitem{densenet}
\BIBentryALTinterwordspacing
G.~Huang, Z.~Liu, and K.~Q. Weinberger, ``Densely connected convolutional
  networks,'' \emph{CoRR}, vol. abs/1608.06993, 2016. [Online]. Available:
  \url{http://arxiv.org/abs/1608.06993}
\BIBentrySTDinterwordspacing

\bibitem{ExFuse}
\BIBentryALTinterwordspacing
Z.~Zhang, X.~Zhang, C.~Peng, D.~Cheng, and J.~Sun, ``Exfuse: Enhancing feature
  fusion for semantic segmentation,'' \emph{CoRR}, vol. abs/1804.03821, 2018.
  [Online]. Available: \url{http://arxiv.org/abs/1804.03821}
\BIBentrySTDinterwordspacing

\bibitem{effectiveness_receptive_field}
W.~Luo, Y.~Li, R.~Urtasun, and R.~Zemel, ``Understanding the effective
  receptive field in deep convolutional neural networks,'' in \emph{Advances in
  Neural Information Processing Systems}, 2016, pp. 4898--4906.

\bibitem{SPP}
\BIBentryALTinterwordspacing
K.~He, X.~Zhang, S.~Ren, and J.~Sun, ``Spatial pyramid pooling in deep
  convolutional networks for visual recognition,'' \emph{CoRR}, vol.
  abs/1406.4729, 2014. [Online]. Available:
  \url{http://arxiv.org/abs/1406.4729}
\BIBentrySTDinterwordspacing

\bibitem{BLIINDS_II}
M.~A. Saad, A.~C. Bovik, and C.~Charrier, ``Blind image quality assessment: A
  natural scene statistics approach in the dct domain,'' \emph{IEEE
  Transactions on Image Processing}, vol.~21, no.~8, pp. 3339--3352, Aug 2012.

\bibitem{M3}
W.~Xue, X.~Mou, L.~Zhang, A.~C. Bovik, and X.~Feng, ``Blind image quality
  assessment using joint statistics of gradient magnitude and laplacian
  features,'' \emph{IEEE Transactions on Image Processing}, vol.~23, no.~11,
  pp. 4850--4862, Nov 2014.

\bibitem{NFERM}
K.~Gu, G.~Zhai, X.~Yang, and W.~Zhang, ``Using free energy principle for blind
  image quality assessment,'' \emph{IEEE Transactions on Multimedia}, vol.~17,
  no.~1, pp. 50--63, Jan 2015.

\bibitem{NIQE}
A.~Mittal, R.~Soundararajan, and A.~C. Bovik, ``Making a ¡°completely blind¡±
  image quality analyzer,'' \emph{IEEE Signal Processing Letters}, vol.~20,
  no.~3, pp. 209--212, March 2013.

\bibitem{ILNIQE}
L.~Zhang, L.~Zhang, and A.~C. Bovik, ``A feature-enriched completely blind
  image quality evaluator,'' \emph{IEEE Transactions on Image Processing},
  vol.~24, no.~8, pp. 2579--2591, Aug 2015.

\bibitem{QAC}
W.~Xue, L.~Zhang, and X.~Mou, ``Learning without human scores for blind image
  quality assessment,'' in \emph{IEEE Conference on Computer Vision and Pattern
  Recognition}, June 2013, pp. 995--1002.

\bibitem{LPSI}
Q.~Wu, Z.~Wang, and H.~Li, ``A highly efficient method for blind image quality
  assessment,'' in \emph{IEEE International Conference on Image Processing},
  Sept 2015, pp. 339--343.

\bibitem{resnet}
\BIBentryALTinterwordspacing
K.~He, X.~Zhang, S.~Ren, and J.~Sun, ``Deep residual learning for image
  recognition,'' \emph{CoRR}, vol. abs/1512.03385, 2015. [Online]. Available:
  \url{http://arxiv.org/abs/1512.03385}
\BIBentrySTDinterwordspacing

\bibitem{data_augmentation}
\BIBentryALTinterwordspacing
L.~Taylor and G.~Nitschke, ``Improving deep learning using generic data
  augmentation,'' \emph{CoRR}, vol. abs/1708.06020, 2017. [Online]. Available:
  \url{http://arxiv.org/abs/1708.06020}
\BIBentrySTDinterwordspacing

\bibitem{ImageNet}
J.~{Deng}, W.~{Dong}, R.~{Socher}, L.~{Li}, {Kai Li}, and {Li Fei-Fei},
  ``Imagenet: A large-scale hierarchical image database,'' in \emph{IEEE
  Conference on Computer Vision and Pattern Recognition}, June 2009, pp.
  248--255.

\bibitem{kaiming_uniform}
K.~{He}, X.~{Zhang}, S.~{Ren}, and J.~{Sun}, ``Delving deep into rectifiers:
  Surpassing human-level performance on imagenet classification,'' in
  \emph{IEEE International Conference on Computer Vision}, Dec 2015, pp.
  1026--1034.

\bibitem{SGD}
I.~Sutskever, J.~Martens, G.~Dahl, and G.~Hinton, ``On the importance of
  initialization and momentum in deep learning,'' in \emph{International
  Conference on International Conference on Machine Learning}, 2013, pp.
  III--1139--III--1147.

\bibitem{PWRC}
Q.~{Wu}, H.~{Li}, F.~{Meng}, and K.~N. {Ngan}, ``A perceptually weighted rank
  correlation indicator for objective image quality assessment,'' \emph{IEEE
  Transactions on Image Processing}, vol.~27, no.~5, pp. 2499--2513, May 2018.

\bibitem{cross-validation}
M.~W. Browne, ``Cross-validation methods,'' \emph{Journal of Mathematical
  Psychology}, vol.~44, no.~1, pp. 108 -- 132, 2000.

\bibitem{complexity}
\BIBentryALTinterwordspacing
P.~Molchanov, S.~Tyree, T.~Karras, T.~Aila, and J.~Kautz, ``Pruning
  convolutional neural networks for resource efficient transfer learning,''
  \emph{CoRR}, vol. abs/1611.06440, 2016. [Online]. Available:
  \url{http://arxiv.org/abs/1611.06440}
\BIBentrySTDinterwordspacing

\end{thebibliography}
\end{document}